\documentclass[preprint2]{aastex}

\newcommand{\muas}{$\mu{\rm as}$}
\slugcomment{Submitted to PASP}

\shorttitle{Astrometry in Crowded Fields with SIM - I}
\shortauthors{R.\ Sridharan \& R.J.\ Allen}

\begin{document}

\title{{\sc Crowded-Field Astrometry with the \\
Space Interferometry Mission - I \\[0.1in]
Estimating the Single-Measurement Astrometric Bias \\
Arising from Confusion} \\[0.2in]
}

\author{R.\ Sridharan \& Ronald J. Allen}
\affil{Space Telescope Science Institute, 3700 San Martin Drive, 
Baltimore, MD 21218}
\email{sridhar@stsci.edu, rjallen@stsci.edu}

\begin{abstract}
The accuracy of position measurements on stellar targets with the future Space
Interferometry Mission (SIM) will be limited not only by photon noise and by
the properties of the instrument (design, stability, etc.) and the overall
measurement program (observing strategy, reduction methods, etc.), but also by
the presence of other ``confusing'' stars in the field of view (FOV). We use
a simple ``phasor'' model as an aid to understanding the main effects of this
``confusion bias'' in single observations with SIM. This analytic model has been
implemented numerically in a computer code and applied to a selection of typical
SIM target fields drawn from some of the Key Projects already accepted for the
Mission. We expect that less than 1\% of all SIM targets will be vulnerable to
confusion bias; we show that for the present SIM design, confusion may be a
concern if the surface density of field stars exceeds 0.4 star/arcsec$^2$.
We have developed a software tool as an aid to ascertaining the possible
presence of confusion bias in single observations of any arbitrary field. Some
\textit{a priori} knowledge of the locations and spectral energy distributions
of the few brightest stars in the FOV is helpful in establishing the possible
presence of confusion bias, but the information is in general not likely to be
available with sufficient accuracy to permit its removal. We discuss several
ways of reducing the likelihood of confusion bias in crowded fields. Finally,
several limitations of the present semi-analytic approach are reviewed, and
their effects on the present results are estimated. The simple model presented
here provides a good physical understanding of how confusion arises in a single
SIM observation, and has sufficient precision to establish the likelihood
of a bias in most cases. We close this paper with a list of suggestions for
future work on this subject.
\end{abstract}


\section{Introduction}
\label{intro}
The Space Interferometry Mission (SIM\footnote{also currently called
SIM--PlanetQuest}) is being designed by NASA's Jet Propulsion Laboratory to
provide a facility-class instrument for measuring the positions and proper
motions of stars at optical wavelengths with micro-arc-second (\muas) precision.
This represents an improvement by several orders of magnitude over the
precision of all existing astrometric instruments. For faint sources
(V $\gtrsim 15$ mag.), SIM will also be more than a factor 10 better than any
other future planned space mission, and will therefore uniquely permit many
new classes of problems to be addressed. Such problems include: the direct 
measurement (for the first time) of the masses of earth-like planets in orbit
around nearby stars; determining the distances to stars by direct triangulation
over the whole Galaxy and out to the Magellanic Clouds; measuring the transverse
motions of galaxies in the Local Group; and, establishing the shape of the dark
matter distribution in the Galaxy. A description of the instrument and its
current science program is available at the JPL/SIM web site.\footnote{See
{\it http://planetquest.jpl.nasa.gov/SIM/sim\_index.cfm} for
descriptions of the current set of Key Projects. Almost half of the available
5-year Mission time is still unallocated.}
The mission is now at the end of the detailed design phase. After more than 15
years of development, all the major technical questions have been answered.
New devices have been invented in order to provide metrology internal to the
spacecraft at a level of a few tens of picometers, a fraction of the
inter-atomic distance in a molecule of oxygen. The next major step is to begin
construction of the instrument.

SIM will be the second optical interferometer in space devoted to astrometry,
following the Fine Guidance Sensors (FGS) on the Hubble Space Telescope (HST).
However, SIM is a Michelson interferometer using separated collectors, quite
different from the filled-aperture white-light shearing interferometer design
of the FGS. SIM includes three long-baseline interferometers housed on a common
truss, each formed by two $\approx 0.3$ m apertures which compress their light
beams and guide them through delay lines to beam combiners. During operation,
two interferometers are used for precision guiding of the spacecraft, and the
third views the target of interest. Data are then accumulated by tracking
the target until enough photons have been recorded to achieve the particular
science goal.

The precision with which astrometry can be done on a specific stellar target
with SIM will be limited by photon noise, the design and stability of the
instrument, and by the data calibration and processing. We have some control
over the instrument properties, which depend on the specific choices made when
implementing the design in hardware. Also, the operation and calibration of the
instrument can be optimized so as to maximize the achievable
precision. However, there is another source of error which may be present,
and which is largely out of our control; it does not reduce the ultimate
\textit{precision} with which a given \textit{single measurement} can be
made with SIM, but it may reduce the final \textit{accuracy} of that measurement. This source of error arises because of the presence of other stars
in the SIM field of view (FOV) which can ``confuse'' any single observation made
on the target star. The light from these extraneous stars perturbs the measurement, so that the measured target position can differ from the true
position. The difference is a \textit{bias} which can reach a level of many 
times the single-measurement precision estimated from the instrument parameters
alone. It is this \textit{single-measurement confusion bias} which concerns us
in this paper.

It is important to emphasize that the final \textit{accuracy} of the astrometric
parameters (position, parallax, and proper motion) determined by SIM for any given target star will be a result of carrying out a complex program of several
single measurements on that target, plus repeated measurements on many other
stars for the determination of calibration and baseline orientation parameters \citep{bod04,miltur03}.
Since we are concerned only with possible bias in a \textit{single
measurement}, the details of the entire observing program are not directly
relevant here; however, it also means that we can not quantify the consequences
of confusion bias to the final accuracy with which the ``end-of-mission''
astrometric parameters can be obtained on any specific target. It is clear that
the effects of single-measurement confusion bias will generally diminish
as more observations are combined. But this also means that projects which
involve only a few observations of a target (e.g.\ ``targets of opportunity'',
single parallax measurements on nearby bright stars, etc.) may have a greater
susceptibility to confusion bias.

Specific aspects of confusion in astrometric measurements with SIM have been
considered by several authors in the recent past. \citet{dalalgriest01} showed
that the characteristic response of SIM's fixed-baseline interferometer as a
function of wavenumber and delay can be used to refine a model of the
distribution of confusing stars in the FOV. This model can then be used to
correct the measured position of the target of interest, and in many cases
where the level of confusion is not too great the astrometric accuracy can
approach the measurement precision.
Dalal and Griest successfully applied their method to models of confused
fields in the LMC in which $\approx 16$ faint stars are scattered
over the FOV around the $\approx 19$ magnitude target star. Photon noise is
included in these models. These authors then go on to consider the additional
complication if one of the stars in the FOV changes brightness owing to a
micro-lensing event, and show that an extension of their fitting algorithm to
include the precision photometry provided by SIM's detectors permits even this
apparently-intractable case to be handled almost as well.
However, their method fails when the angular separation between any pair of
sources in the FOV (as projected on the interferometer baseline) corresponds
to a delay difference of $\lesssim 2$ coherence lengths for the full bandpass
of the detection system. This is a projected separation of 25 milli-arcseconds
(mas). Indeed, this is a \textit{general limit} for SIM observations. Our approach is somewhat simpler than that of Dalal and Griest, and yields
some improvement in the minimum angular separation which can be measured, but
the basic limitation can not be overcome. We will compare our approach to
theirs in more detail in a future paper.

\citet{rajbokall01} also considered a number of specific cases of confusion on
SIM astrometry. These authors introduced a graphical analogy using phasors as an
aid to understanding how errors in the target position arise from confusing
sources in the FOV. Typical target fields were constructed on a simulated image
with grid spacing of 5 mas, and the amplitude and phase of the fringe which
would be measured with SIM for a given wavelength on that image field was
computed with a Fourier transform. Diffraction effects at the edges of the
(presumed $\approx 1''$ square) SIM FOV were included in constructing the model
image, and vector averaging of the individual (narrow) SIM wavelength channels
was used to simulate the 1-dimensional apodization of the fringes over the FOV
caused by the decreasing coherence of the fringes as the bandwidth increases.
\citet{rajbokall01} were particularly interested in modeling the effects of
mispointing of the FOV in subsequent visits to the same target field when
target proper motions were being measured; in that case the actual distribution
of field stars changes as some disappear from one side of the FOV and others
appear at the other side. They included photon noise, and also addressed the
issue of how the size of the FOV defined by the field stop influenced the level
of confusion. Their source field models were constructed to simulate SIM
observations of the position and proper motion of target stars in M31, the LMC,
and the Galactic bulge. They concluded that the confusion-induced errors in
position can often be significant (several times the measurement precision)
for faint target stars but the proper motion errors are likely to be small. 
The errors are, as expected, smaller for wider measurement bands.

In one other confusion-related study, \citet{takvellin05} considered the
effect of circumstellar disks on the measurement of stellar wobble during
observations aimed at detecting extra-solar planets. Their models showed
that neither the motion of the disk mass center nor the contamination
by disk light is a serious threat to detecting planets around pre-main
sequence stars; the basic reason for the insensitivity of these observations
to confusion from circumstellar disks is that interferometers tend to resolve
such an extended source, reducing its influence on the astrometry of the parent
star.

The studies summarized above have shown that confusion poses limits
to the accuracy of any single SIM measurement, and have therefore succeeded
in raising our awareness of this problem.  However, the detailed design
of SIM has changed since those studies were done, and many of the changes
will affect on the modeling results. The size of the collector and
its central obscuration, the entrance aperture field stop defining the  
geometrical FOV, the transmission efficiency of the optics, the fringe
disperser design (which defines the bandwidth and central wavelength of the
spectral channels), and the QE and spectral response of the detector are now
all much better defined. Furthermore, previous studies have focused
on specific science programs which were already suspected to be pushing the
capabilities of the instrument; they have not provided us with any
general ``tools'' for understanding and recognizing confusion, or for dealing
with it. Previous studies have also often taken a statistical approach which
is less suitable for answering direct questions about specific fields, such as:
is a SIM observation of this particular target embedded in that particular field
of stars likely to be confused? And, can the observation be done in
such a way so as to reduce the confusion bias?  What \textit{a priori}
information about the target and the field would help? And, if the
observation has already been taken, can we identify the effects of confusion in
the data? These questions have provided the motivation for the work described in
this paper.

This paper is organized as follows. In the next section, we summarize the basic
Michelson interferometer response as it applies to SIM. We then recall the
phasor model introduced by \citet{rajbokall01} and elaborate upon it as a tool
for understanding the behavior of confusion in SIM astrometry. Using this
analytic model, together with updated knowledge of SIM's instrument parameters,
we have constructed a simulation code for evaluating the likelihood of confusion
bias in any specific field; details are presented in Section~\ref{simu}.
In Section~\ref{limit_values}, we present single measurement confusion bias as a function of  
magnitude difference and projected separation of an additional star present
within the SIM FOV.
In Section~\ref{applications}, we apply this semi-analytic model to a number of
target fields drawn from the Key Projects which have already been chosen for the
initial SIM science program. From this experience we then consider how the
single-measurement confusion bias might be reduced through the addition of other
information. The most useful additions appear to be knowledge of the approximate
locations and spectral energy distributions (SEDs) of the target and of the most
troublesome confusing stars in the SIM FOV. Finally, in Section
\ref{limitations}, we describe the limitations of our current approach. These
limitations are primarily related to the simplified model of the focal plane of
SIM  which we have used here. In particular, in this
paper, we have not modeled the detailed mechanism by which the spectral
dispersion is implemented,\footnote{A thin prism disperser turns SIM into an
objective prism spectrograph on the CCD detector.} nor have we considered
the pixellation of the focal plane by the CCD detector. We have explored these
points with the aid of a more elaborate model that includes these effects, and find that the biases estimated using this more elaborate model differ only by
small amounts from those provided by the approach described here.\footnote
{However, consideration of this more sophisticated model does suggest
additional ways to reduce any confusion bias.} We have therefore chosen to
present the main issues relevant to SIM confusion with a minimum of
complication, and leave the discussion of the more elaborate instrument model
to a future paper.

Binary stars will be an important class of targets for SIM, and are the
topic of one of the major Key Projects. In these cases, the two stars are in
a bound orbit and are physically close to each other. Typical binaries to be
studied with SIM will have separations from about a few mas to 1000 mas and 
orbital periods from a few days to several years.  Stars in crowded
fields can occasionally mimic the effects of binaries if their projected
separations become small for particular baseline orientations, but the effects
of confusion from such ``apparent'' binaries can be reduced (or even eliminated)
by rotating the interferometer baseline and repeating the observation. However,
for ``real'' binaries, rotation of the baseline is an integral part of the
measurement process. The goal of the binary observation is to obtain the
characteristics of the orbit, and this is done by measuring the positions of
the components for a number of baseline orientations. These targets
are sufficiently specialized that we have removed them from the list of
crowded-field problems treated in this paper. Such observations treat binaries
as ``signal", whereas here we treat them as ``noise".  A discussion of astrometry on
binary stars with SIM is planned for a future publication.

\section{Astrometry with SIM}

Historically, the angular separation between two stars on the sky has been
determined using telescopes of various sizes mounted on mechanical structures
including large metal segments with regular rulings on their edges, or a
combination of such devices with measurements of transit times. An astrometric
interferometer converts the problem from a measurement of
\textit{angular increments} to a measurement of a \textit{distance interval}
which can be determined very precisely using laser metrology inside the
instrument. Thanks to the absence of atmospheric
instabilities in the space environment where SIM will operate, an additional
level of precision worth several orders of magnitude can be added by
directly measuring the \textit{phase} of the fringe pattern using target
photons. The main features of how these measurements are made with SIM will
now be described.

\subsection{The SIM interferometer}

The response pattern $P$ of a Michelson interferometer as applied to astrometry
and imaging in astronomy can be written, under quasi-monochromatic conditions, 
as follows:
\begin{eqnarray} \label{eqn:SIMresponse}
\lefteqn{P(\delta, \theta, \overline{\lambda}) = }\\
 &    P_{0} \{ 1 + A \sin (\frac{2 \pi}{\overline{\lambda}}
 [\delta - B \theta ] ) \}. \nonumber
\end{eqnarray}
where, $P_0$ is the total light collected by the two collectors, $A$ is the
fringe amplitude, $\delta$ is the internal path delay, $B$ is the baseline,
$\theta$ is the angle on the sky, and $\overline{\lambda}$ 
is the mean wavelength. See Appendix \ref{appsec:response} for the derivation
of this equation. For convenience we often use the mean wavenumber
$\overline{k} \equiv 1/\overline{\lambda}$.

\begin{figure}
\epsscale{0.8}
\plotone{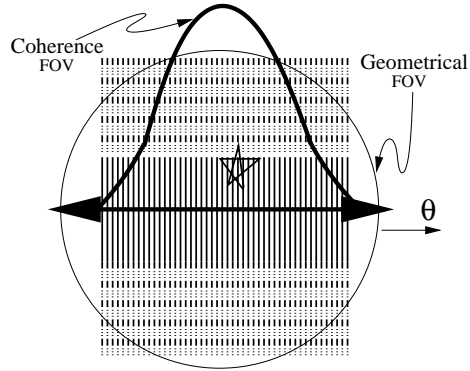}
\caption {The response of the interferometer as a function of the angle on the
sky, for a fixed internal delay $\delta$ and wavelength $\lambda$. A dense
pattern of 1-D fringes  showing the response of the instrument as a function of
position is here projected onto a 2-D FOV on the sky. The interferometer
baseline orientation is indicated by the thick double-headed horizontal arrow.
The amplitude of the fringes is apodized by a 1-D ``coherence FOV'' function
which depends chiefly on the bandwidth of the channel, as described in
Appendix \ref{appsec:response}. \label{fig:FOV}
}
\end{figure}

\begin{figure}
\epsscale{0.8}
\plotone{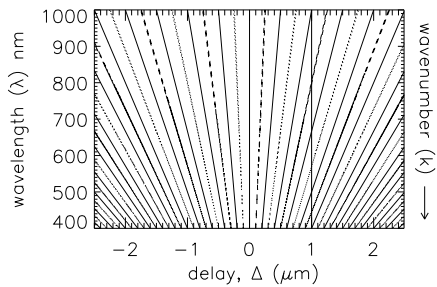} 
\caption {The ``fan diagram''; a contour plot of the Michelson
interferometer response as a function of total path delay
$\Delta = \delta - B\theta$ and wavelength. 
The fringes appear to `fan out' at longer wavelengths, hence the name. 
The possible location of a target star is shown at $\Delta = +1 \mu$.
\label{fig:FANdiagram}
}
\end{figure}

The response of the interferometer as a function of $\theta$
projected onto the two-dimensional sky is sketched in Figure~\ref{fig:FOV}.
Figure \ref {fig:FANdiagram} shows the same interferometer response, now
as a function of the total optical path difference
$\Delta = \delta - B\theta$ and $\lambda$. This Figure is also known as
the ``fan diagram". 
In this diagram, stellar targets appear at specific values of total delay
$\Delta_i$
along the $x$ axis corresponding to their actual positions $\theta_i$ in the
sky and the setting of the internal delay $\delta$,
and so can be represented as a set of
vertical lines. An example is shown in Figure~\ref{fig:FANdiagram} for a single
target at $\Delta$ = +1$\mu$.  Moving vertically in this diagram from shorter
to longer wavelengths at a given delay $\Delta$, we see that the fringe phase changes linearly if the delay is close to the location of an isolated target,
and will jump by $\pm 2 \pi$ radians if the delay is large.
If the delay setting corresponds to
the actual location of the target, the fringe phase will be a constant; this is
the position at which the total optical path delay difference between
the two arms of the interferometer is zero.  It is this particular 
property of the fringes which enables us to identify the position of the 
target precisely.

\subsection{How SIM works}
\label{how_sim_works}

\begin{figure}
\epsscale{0.8}
\plotone{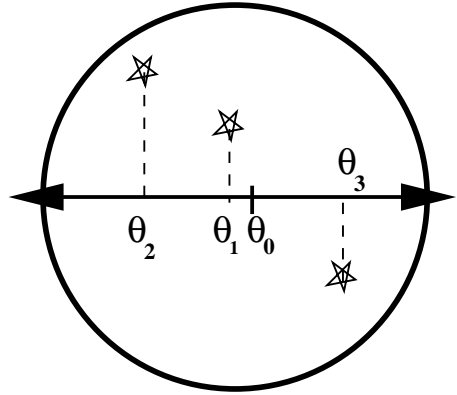}
\caption {
A potentially-confused FOV consisting of a target star at $\theta_1$
and two field stars at
$\theta_2$ and $\theta_3$. $\theta_0$ corresponds to the current setting of the
internal ``coarse'' delay $\delta_c$. Scanning the ``fine'' delay $\delta_f$
and fringe fitting will lead to the identification of a position which can be
different from the delay offset corresponding to the ``true'' location of the
target. Thus the presence of field stars can introduce a bias in the
measurement of the target star's position. \label{fig:model2}
}
\end{figure}

Consider a single target star located at $\theta_1$ near the center of the FOV
as shown in Figure \ref{fig:model2} (ignore the other stars located at
$\theta_2$ and $\theta_3$ for the time being).
It is now convenient to consider the internal delay to be  made up of two parts, a ``coarse'' part and a ``fine'' part, such that $\delta = \delta_c + \delta_f$.
The measurement begins by setting $\delta_f = 0$ and adjusting $\delta_c$ until
fringes are found. The bandwidth is then progressively widened (by further
binning the channels) and the coarse delay further adjusted so as to maximize
the fringe contrast. A limit will reached when the delay difference
$\delta_c - \delta_1$ is approximately within the coherence length appropriate
for the maximum available instrument bandwidth. Defining $\delta_0 \equiv
B \theta_0$ as this ``final'' setting of $\delta_c$, the goal of the subsequent
steps in the measurement process is to measure $\delta_1 - \delta_0 \equiv
B (\theta_1 - \theta_0)$ by scanning the fine delay in a series of small
steps to a maximum of $\approx \pm 1 \lambda$ around zero, and recording the
fringes which appear in each of $n$ narrow-band channels. The response function
of equation \ref{eqn:SIMresponse} now reads:
\begin{eqnarray} \label{eqn:finalSIMresponse}
\lefteqn{P(\delta_f, \phi_n, \overline{k_n}) = } \\
  &  P_{0}(n) \{ 1 + 
       A_n \sin [ 2 \pi \overline{k_n}\delta_f + \phi_n ) ] \} \nonumber
\end{eqnarray}
where $\phi_n = 2 \pi \overline{k_n} B (\theta_0 - \theta_1)$. A fit is then
done to the data in each channel, yielding the fringe parameters $P_0(n),\,
A_n,\, \mbox{and } \phi_n$.

\begin{figure}
\epsscale{0.8}
\plotone{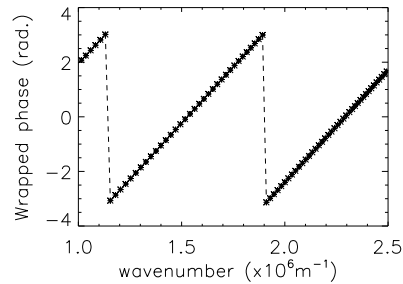}
\caption {The phase spectrum, showing the fringe phase $\phi$ vs.\ wavenumber. Note the ``phase-wrap'' jumps. \label{fig:phase_vs_wavenumber}
}
\end{figure}

\begin{figure}[htb]
\epsscale{0.8}
\plotone{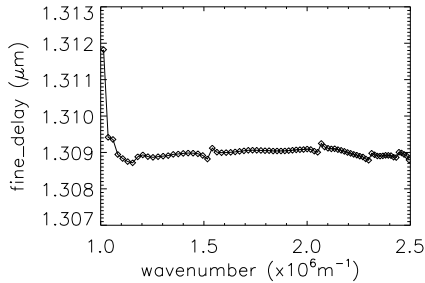}
\caption {Delay spectrum estimated from the unwrapped fringe phases of Figure
\ref{fig:phase_vs_wavenumber} as described in the text.
The average delay is 1309.02~nm, and corresponds to an offset of 30000.58 \muas.
The ``glitches'' on this figure are a result of simplifications we have made in
our numerical model for the dispersion of the fringes with wavelength.
\label{fig:delay_vs_wavenumber}
}
\end{figure}

A plot of the resulting ``phase spectrum'' $\phi_n$ vs.\ the mean wavenumber
$\overline{k_n}$ is shown in Figure~\ref{fig:phase_vs_wavenumber}. This plot
will show jumps of the fringe phase if the delay offset $\delta_1 - \delta_0$
is larger than a typical fringe period. Taking the phase at one of the channels
as a reference, the phases of the other channels can be ``unwrapped'' by adding
or subtracting $2 \pi$ when a discontinuity is encountered.  A plot of the
``unwrapped'' phases will be a straight line parallel to the $x$ axis which we
call the ``delay spectrum'', as shown in Figure~\ref{fig:delay_vs_wavenumber}.
The final delay offset (and hence the angular position $\theta_1$) can be
computed from a simple average of the data in this Figure, although several
other possibilities exist \citep{milbas02}.  

We have glossed over a large number of calibration issues, including the
precise measurement of delay increments along the internal delay line, the precise length of the baseline, the calibration of the relation $\delta_0 \Leftrightarrow \theta_0$, etc. In addition, the measurement we have
obtained is an angle with respect to the direction of the interferometer
baseline, but that direction is not yet known in any external reference
frame such as equatorial coordinates. Precise determination of the baseline
vectors used for all SIM observations is clearly an extremely important part
of the whole astrometric program. These calibration and baseline
determination problems are a major part of the SIM project.
However, confusion bias acts as a perturbation on a single SIM observation, so it is possible to discuss the
origin and nature of confusion bias in single measurements without a detailed
consideration of the entire astrometric program of SIM.

\subsection{Confusion bias in SIM measurements}

The operation of SIM as described in the previous section makes the assumption
that the FOV contains only a single target star, and this is expected to be the
case for more than 99\% of the fields to be observed. However, the real sky may
occasionally contain other stars in the foreground, background, or even
associated with, the target star, as shown in the sketch of Figure
\ref{fig:model2}. Since the fringe patterns of all the stars in the FOV are the same in the same wavelength channel, the resulting fringe pattern for several
stars together will mimic that of a single star, but with a different amplitude
and phase. If the bandwidth is wide, and the (baseline-projected) positions of
the confusing stars are well away from the target star, their influence will
be attenuated. Nevertheless, it is clear that the presence of such extraneous
``field stars'' in the FOV can lead to a bias in the measured fringe phase, and
therefore to a bias in the measured position of the target. The final
measurement \textit{precision} may not be affected, but the final
\textit{accuracy} of the target star position will be reduced.

\subsection{Visualizing confusion}
\label{subsec:visconf}

\begin{figure}[ht]
\epsscale{1.0}
\plotone{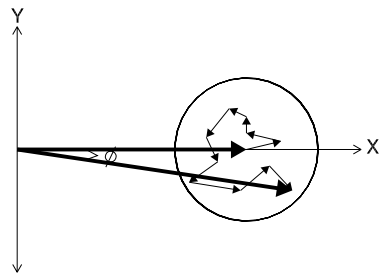}
\caption{
Phasor diagram at a specific wavenumber $k$ showing the vector addition of
the complex visibilities of a bright target star, located at the origin of the
coordinate system, with the visibilities of 10 fainter stars, located elsewhere
in the FOV.
\label{fig:phasor_diagram}
}
\end{figure}

One can visualize the nature of confusion with the help of Argand diagrams
from the theory of complex variables \citep[e.g.][]{phi61} since, 
after subtraction of the total power term, the sine-wave fringe pattern of
Equation~\ref{eqn:SIMresponse} produced by an interferometer observing 
a single un-confused target star maps one-to-one with a vector in the 
complex plane having \textit{modulus} equal to the fringe amplitude and \textit
{argument} equal to the fringe phase. Such vectors are called \textit
{phasors}\citep[e.g.][]{gas78,hec02}, and the results of adding many
fringes together from many stars in a crowded field are easily understood in
terms of a vector sum of these phasors as shown in Figure
\ref{fig:phasor_diagram}, adapted from Figure 1 of \cite{rajbokall01}.
The bright target is assumed to be located at the origin of the (local)
coordinate system, and therefore has zero fringe phase; it is represented
by a vector along the real
axis with some amplitude. The field stars have non-zero fringe phases; they are
represented by shorter phasors with relative magnitudes and directions depending
on  their locations in the field with respect to the target. The field stars
form a ``noise cloud'' at the tip of the strong phasor representing the
astrometric target of interest, and their combined effect is to alter the final
fringe amplitude and phase. The shift $\phi$ of the fringe phase from the
true position of the target phasor is the bias owing to confusion. Since the
phases of the confusing sources change with observing wavelength, and at a rate
which depends on their (projected) displacement from the nominal field center,
the small phasors all rotate at different rates and directions as the observing
wavelength changes, so that the net confusion bias $\phi = \phi(\lambda)$ of the
 target of interest changes with the wavelength of the observations. It is now
obvious that the effects of confusion will depend on the spectral energy
distributions (SEDs) of the target and of all the confusing sources. 

\section{Simulating SIM's response}
\label{simu}

In order to proceed with a simulation we need to know the expected structure
of the distribution of light in (and surrounding) SIM's FOV, and the response
of the instrument to this distribution.

\subsection{Source models}

In general, we assume that source distributions can be approximated as a set of
unresolved sources. We ignore the presence of extended emission, which will in
any case contribute little to the net fringe amplitude (but may increase the
photon noise on the detector). Each of these sources is represented by a delta
function with some amplitude and some 2-D location in the FOV. The spatial
distribution is either idealized from existing  images of the target area,
e.g.\ images from the cameras of the \textit{Hubble Space Telescope} (HST), or
defined \textit{ad hoc} with an educated guess. A model consists of a set of
x, y positions within (and just outside of) the FOV, along with the brightnesses
and the spectral energy distributions (SEDs) of the sources at each position. If
the actual SEDs are not available, model SEDs are either approximated or
obtained from spectral libraries.\footnote{ From the web site\\ {\it http://garnet.stsci.edu/STIS/stis$\_$models.html$\#$models}\\
when spectral types are known and from the file {\it{kurucz$\_$orig.fits}} available at the 
web site\\ {\it http://dae45.iaa.csic.es:8080/$\sim$jmaiz/software/chorizos\linebreak[1]/chorizos.html} when gravity, effective temperature and metallicity are known.}

\subsection{Instrument models}

We provide here a high-level description of the SIM instrument model we have
adopted in order to determine the total power, fringe amplitude and fringe
phase of a given source distribution. The numerical details and relevant 
figures are collected in Appendix~\ref{appsec:power_estimation}.

\subsubsection{Incident photon fluxes}

The number of photons detected from each star within the FOV is estimated
as follows:

\subsubsubsection{Channel response}

We estimate the mean wavelength and bandpass of all 80 channels of SIM
from the knowledge of dispersion (wavelength vs.\ position at the focal plane)
and detector pixel size. We assume that each pixel corresponds to one channel,
and that the total energy of all the stars within the FOV is collected in a single pixel. This is an idealization, but we defer details of the dispersion
and the pixellation in the focal plane to a later paper where a more complete
instrument model is presented. As we shall discuss further in \S
\ref{limitations}, these additional complications have only small
effects on our results, and come at the expense of considerable
additional complexity. 

\subsubsubsection{Apodization through the field-stop}

For each star in the FOV, the effects of diffraction through the entrance aperture and
apodization by the field stop are modeled as the convolution of a circular
stop with the (wavelength-dependent) diffraction pattern of SIM's entrance
aperture including the effects of the central obscuration (created to
accommodate the internal laser metrology system). This gives the aperture
transmission factor (a fractional number as defined here)
which, when multiplied by the total power, gives the fraction of the
total light from the star that falls within the field stop.
See Appendix \ref{appsec:power_estimation} for the details.

\subsubsubsection{Throughput}

The optical train of SIM involves many reflections, each with some loss which
depends on the wavelength, and ultimately the light is imaged on a detector 
which itself has some wavelength response. These various reflectivities have
been multiplied with the net CCD detector quantum efficiency in order to obtain
an overall wavelength-dependent ``throughput'' of the optical system.  
We have fitted this function to a polynomial for convenience in numerical
computations; the coefficients are listed in Appendix \ref{appsec:params}.
The SEDs, originally available for a mag 10 star, are modified to account 
for the actual magnitude of the stars and are multiplied by twice the collecting
area of a single aperture, the aperture transmission factor, and the throughput, 
integrated over the channel bandpass, and expressed as photons/sec/channel.  

\subsubsection{Fringe amplitude for each star}

The visibility amplitudes of the confusing stars are attenuated by the
bandpass function (cf.\ Equation \ref{eqn:finalresponse}) according to their
angular distances from the position of the target measured parallel to the baseline, and the wavenumber and coherence length for the specific channel.

\subsubsection{Fringe phase for each star}

The visibility phases $\phi_f$ of the field stars depend on their projected
angular distances $\theta_f$ and are given by $\phi_f = 2 \pi \overline{k}
B\sin(\rho \cos(PA-\psi)) = 2\pi \overline{k} B \theta_f$, where 
$\rho$ is the radial distance of the field star measured in
radians, $PA$ is the position angle of the field star, and 
$\psi$ is the position angle of the baseline orientation.
These quantities are defined in a Cartesian coordinate system with origin
at the center of the FOV, and with the $X$ and $Y$ axes oriented
along the directions of Right Ascension and Declination. 

\subsection{Resultant fringe models}

The resultant fringe is the sum of fringes produced by the target and each
of the field stars. In the notation introduced for equation
\ref{eqn:finalSIMresponse}, this total fringe can be written as:
\begin{eqnarray}
\lefteqn{P(\delta_f,\phi_n, \overline{k_n}) =} \nonumber \\
 & \sum_{j=1}^{N} P_0^j(n) \{ 1+A_n^j \sin[2\pi \overline{k_n}\delta_f
 +\phi_n^j] \}, 
\label{eqn:comb1}
\end{eqnarray}
where the summation is over all the $N$ (target + field) stars within
(and just outside) the FOV.
In this quasi-monochromatic approximation, the fringes contributed by each field
star will all have the same period, only their amplitudes and phases will be
different. In that case, the final fringe will also be a pure sinusoid, and we
can write it as
\begin{eqnarray}
\lefteqn{P(\delta_f,\phi_n, \overline{k_n}) =} \nonumber \\
& P_n \{ 1+V_n \sin[2\pi \overline{k_n}\delta_f +\phi_n] \},
\label{eqn:resmod}
\end{eqnarray}
where the parameters are related to those of the field star fringes by
\begin{eqnarray}
P_n &=& \sum_{j=1}^{N} P_0^j(n),   \nonumber \\
P_nV_n \cos\phi_n &=& \sum_{j=1}^{N}P_0^j(n) A_n^j \cos\phi_n^j, \nonumber \\
P_nV_n \sin\phi_n &=& \sum_{j=1}^{N}P_0^j(n) A_n^j \sin\phi_n^j. 
\end{eqnarray}
$P_nV_n\cos\phi_n$ and $P_nV_n\sin\phi_n$ are respectively the cosine and sine
components of the resultant fringe in each of the $n$ narrow band channels.
As assumed in Figure~\ref{fig:phasor_diagram} (and without loss of generality), we take
the target to be located at the center of the FOV, so that $\theta_1 = 0$;
its fringe phase $\phi_n^1$ is therefore zero in all $n$ channels, and the
resultant phase $\phi_n$ is the confusion bias arising from the presence of the
field stars.

Note that $P_0(n), A_n,$ and $\phi_n$ are all functions of the nominal mean
wavelength of the channel, the instrument parameters (as described
earlier), and the intrinsic parameters of the stars such as their effective
surface temperature $T_{eff}$, surface gravity log $g$, metallicity
$\log [Fe/H]$, apparent magnitude $m_{\bar\lambda}$, and location in the FOV.

\subsubsection{Recognizing confusion}

\begin{figure}
\epsscale{0.8}
\plotone{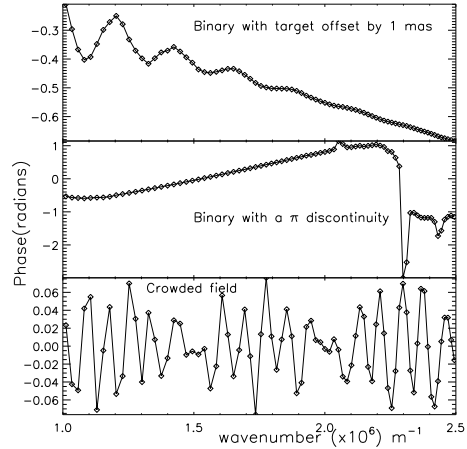}
\caption {Top: Phase spectrum to be expected in the case of a 
binary consisting of a mag 10 A star and a mag 15 M dwarf, separated
by 100 mas along the direction of the baseline orientation.
The resultant phase is a combination of a linear and an oscillatory function.
Middle: A phase jump appears when the binary is resolved at
or close to one of the channels. Bottom: Phase spectrum to be expected in a
crowded field. The resultant phase varies erratically with the wave-number. 
The magnitude of the phases and the resulting distribution across the 
channels determine the severity of the confusion bias.
\label{fig:phasespec}}
\end{figure}

In each FOV we analyze, the model for SIM described above provides us with
80 measurements of the confusion bias $\phi_n$. If the brightness
contrast between the target and the fields stars is large, we expect the
channel-to-channel variation to be small and the resultant phase to be almost
a linear function of channel number, as described in \S \ref{how_sim_works},
with the slope becoming zero if the target is located precisely at the origin of 
the adopted coordinate system in the FOV.
For instance, the top panel in Figure \ref{fig:phasespec} shows the phase
spectrum for such a case; the target is a 10th mag A star, offset from the
FOV coordinate origin by 1 mas, and is confused by a 15th mag M dwarf offset
by a (projected) angle of 100 mas. The linear part of the graph comes from the
1 mas offset of the A star, while the oscillatory part arises because of the
presence of the faint M star. The latter becomes relatively more important
at longer wavelengths, hence the amplitude of the phase ``ripples'' grows
at smaller values of wavenumber. The middle panel shows another example of a binary, this time exhibiting a phase jump at short wavelengths indicating a
position offset of order 10 mas. The bottom
panel of Figure \ref{fig:phasespec} shows the phase spectrum for a crowded
field; the fringe phase in this case varies erratically with channel number, but
remains small.
\begin{figure}
\epsscale{0.8}
\plotone{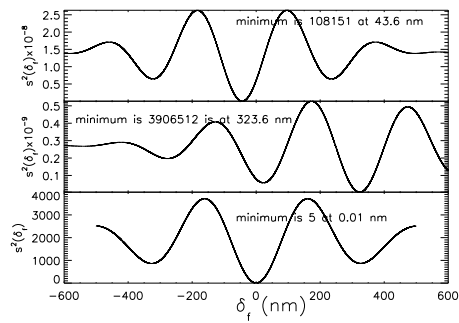}
\caption {The confusion estimator $s^2$ as a function of fine-delay for the three cases
shown in Figure~\ref{fig:phasespec}.  The minimum value of $s^2$ and the corresponding fine-delay
values are indicated for each case. There is some correlation between the amount of deviation of $s^2$ from zero 
to the amount of confusion bias; the exact nature of this relation is not investigated here.}
\end{figure}

\subsection{A confusion estimator}

Phase spectrum diagrams such as those just described (and the related delay
spectrum described in \S \ref{how_sim_works}) are useful indications of the
possible presence of a confusion bias in fields with relatively bright stars,
but their varied character makes it less obvious how to combine them into a
single measure which might be of value for fields of fainter stars. In a real
observation with SIM, one answer to dealing with faint targets is to enlarge
the bandwidth by binning channels on board the spacecraft before computing the
fringe parameters and before transfer of the data to ground. On-board binning
can be accomplished by adding the charge from several pixels as the CCD detector
is read out. This has the advantage of reducing the relative contribution of
read-out noise in the electronics; the disadvantage is that information on the
spectral variations of the fringes is no longer available, making it impossible
to even recognize a confusion bias in a single observation.\footnote{Averaging
data in this way may also mask phase wraps even for unconfused targets.} 
A more thorough discussion of these issues is beyond the scope of this paper,
but may be a useful topic for further work.

In the framework of our simulations, we have developed an estimator for using
all 80 channels of data based on summing the squared fringes in each channel
after subtraction of the DC term. This quantity will vary as the fine delay
$\delta_f$ is scanned; minimizing this squared difference leads to a unique
estimate of the mean astrometric delay, and hence the confusion bias. If the
value of the estimator is zero at that delay to within the noise, the target is
not significantly confused at that signal-to-noise level; non-zero values
are an indication that the measurement is suffering from confusion. The
estimator is written as
\begin{eqnarray}
s^2(\delta_f) \equiv \sum_n \vert P(\delta_f,\xi_n,\overline{k_n})-P_n \vert ^2 \nonumber \\
  = \sum_n \frac{P_n^2V_n^2}{2}\{1- \cos [4\pi\overline{k_n}(\delta_f+\xi_n)]\}
\label{eqn:s_square_definition}
\end{eqnarray}
where the delay spectrum is $\xi_n = \phi_n/2\pi \overline{k_n}$ (cf.\ \S
\ref{how_sim_works} and Figure \ref{fig:delay_vs_wavenumber}), and the other
symbols have been defined previously in connection with equation
\ref{eqn:resmod}. As discussed in \S \ref{how_sim_works}, $\xi_n$ is independent
of the channel number $n$ in the case of isolated targets (after adjustment for
possible phase-wrapping). Thus, the quantity $\delta_f+\xi_n$ on the right side
of Equation~\ref{eqn:s_square_definition} can be considered as an effective
path-length delay of the target, and setting it to zero leads to the precise determination of the delay value $\delta_f^{\prime}$ at which $s^2$ becomes
zero. This is therefore the delay offset of the target from the coarse delay position. Note also that at this value of delay, the fringes disappear, and the
response pattern (Equation~\ref{eqn:resmod}) becomes the SED of the target
modulated by the throughput of the instrument. This discussion also
shows that minimizing the estimator $s^2$ of equation
\ref{eqn:s_square_definition} provides the correct astrometric delay for an
isolated target, and that the value of the estimator $s^2$ at that delay
is exactly zero.

From the discussion above it is also clear that, in the case of a crowded field,
$\xi_n$ varies with channel number and there is no single value of $\delta_f$
which will make each term of the sum in equation \ref{eqn:s_square_definition}
equal to zero. The minimum value of $s^2$ will therefore be a nonzero positive
number, so we will know there is confusion present. But the delay corresponding
to that minimum is the estimator we will use for the target delay.\footnote{To
what extent this is an optimum estimator is presently uncertain.} In our
simulations we know the target position, so we can calculate the true value
of the target delay. We are therefore \textit{defining} the confusion bias in
the simulations to follow as the difference between the known delay of the
target and the delay provided by minimizing the $s^2$ estimator.

It should be emphasized that even though we have defined the confusion estimator
$s^2$ in the context of these simulations, it would be possible to compute
$s^2$ from real SIM data using the values of fringe parameters in each channel
provided by the observations. A fictitious ``fine delay'' could be created
and $s^2$ minimized as usual. To start with, $s^2$ could be estimated
for targets known to be isolated and the noise floor could be determined. Any
non-zero value of $s^2$ for other targets can then be used as an indication
of confusion.

\section{Limiting values of confusion bias}
\label{limit_values}

\begin{figure*}
\epsscale{1.6}  
\plotone{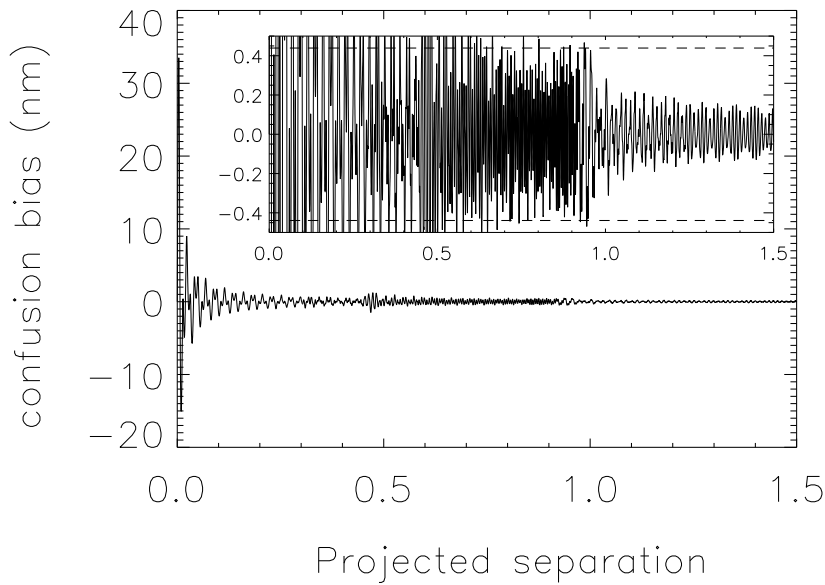}
\caption{Confusion-induced bias in the astrometric delay
of a target star as a function of the projected separation of
the 1 magnitude fainter field star, when both stars have
identical SEDs. The inset indicates a magnified view of a portion
of the plot. The dotted line in the inset indicates the delay bias
of 0.439 nm, which will correspond to a 10 \muas \ position uncertainty
in an single observation. The critical projected distance beyond
which the bias in the delay is less than 0.439 nm is 0.94 arcsec.}
\label{fig:con_vs_proj_sep}
\end{figure*}

\begin{figure*}
\epsscale{2.4}
\plottwo{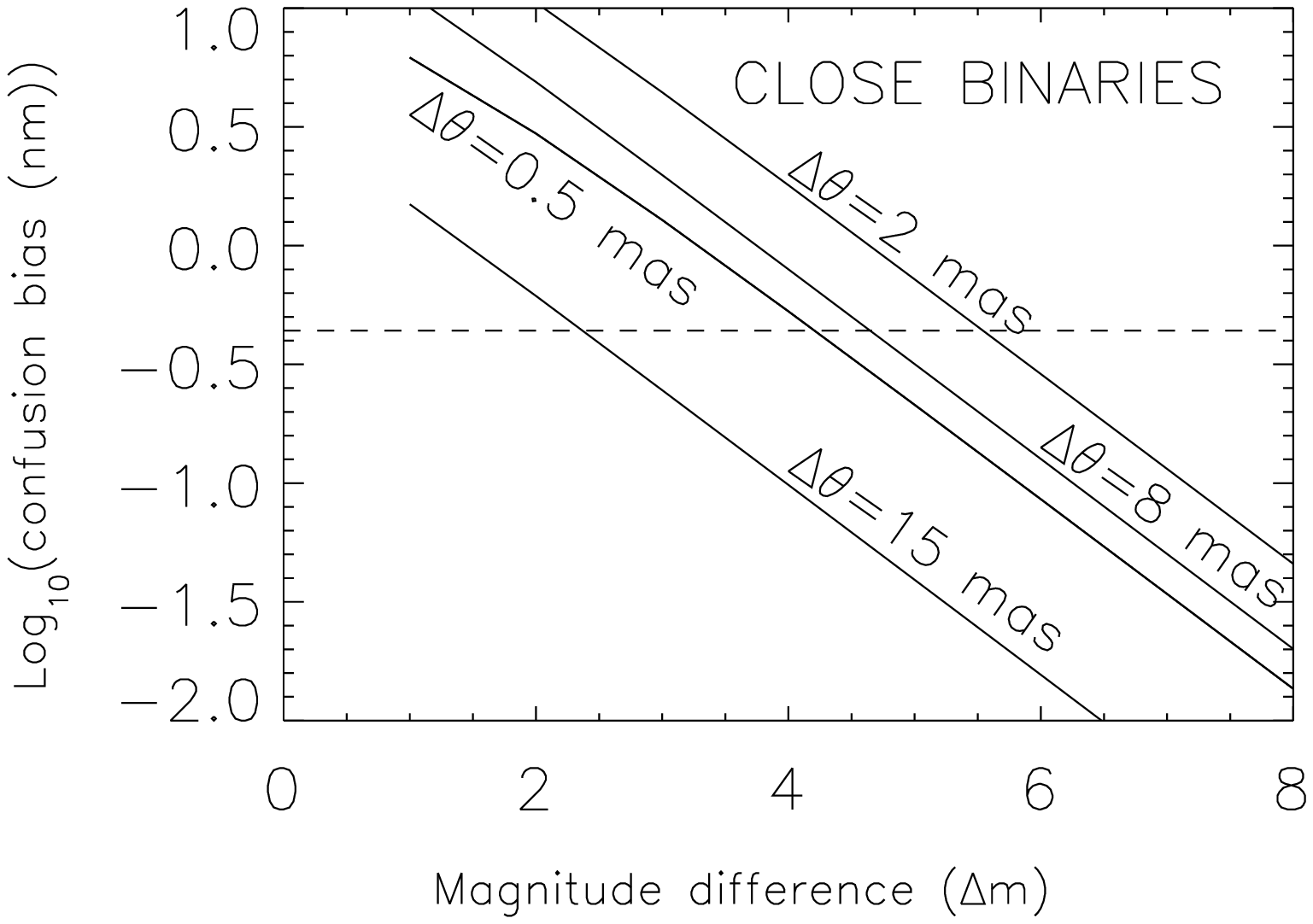}{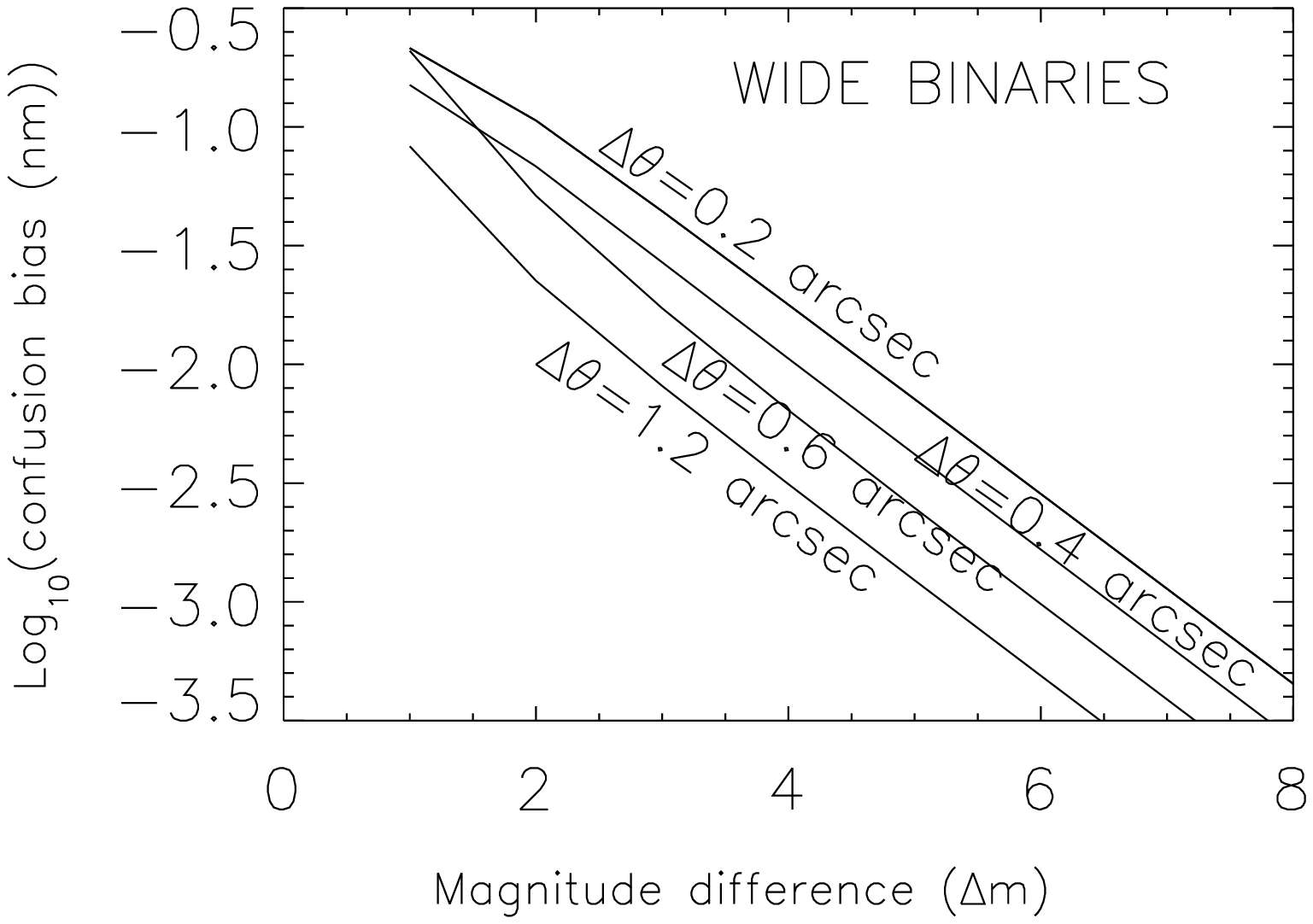}
\caption{Confusion-induced bias in the astrometric
delay of a target star as a function of the magnitude difference between
the target and a fainter field star, both having identical SEDs. The
biases for both close and wide binaries are plotted for a few
arbitrarily-chosen projected separations.
}
\label{fig:con_vs_mag_dif}
\end{figure*}

It is useful to have an estimate of the ``worst case'' values of confusion
bias to be expected in a single SIM observation. While it is possible to
imagine that the phasors for a specific distribution of field stars could
all add up ``in phase'' to produce a very large bias (cf.\ Figure
\ref{fig:phasor_diagram}), this bias would be much reduced in the neighboring
channels; indeed, if there are many field stars, the effects will be to simply
raise the ``noise level'' of the average astrometric delay measurement on the
target. More serious will be those cases where only one or two relatively
bright field stars are present. As an extreme example, we compute the confusion
bias introduced by a single field star, as a function its (projected) angular
distance from the target and the ratio of its brightness to that of the target.
The SEDs of the field and target stars are assumed to be the same. At small
separations, this bias will oscillate with a large amplitude and a period
characteristic of the mean wavelength of the full passband. At larger separations
the amplitude of the oscillations will fall owing to bandwidth apodization
according to equation \ref{eqn:finalresponse}. Figure~\ref{fig:con_vs_proj_sep}
shows the confusion bias (computed using the $s^2$ estimator as described in the
previous section) as a function of projected separation for a field star with
$\Delta$m = +1. For SIM with 9.0~m (projected) baseline, a delay bias of 0.44~nm 
corresponds to an angular position bias
of 10~\muas, which is approximately the expected single-measurement precision,
so this value of delay bias is in some sense a ``critical'' value.
This critical value is reached at a projected separation of $0.94''$ for
a +1 mag field star. The critical projected separations for fainter field
stars with $\Delta$m = 2, 3, 4, and 5 mag are 0.17\arcsec, 0.07\arcsec,
0.024\arcsec, and  0.005\arcsec\ respectively. For instance if a field star
is fainter by 3 magnitudes, it is not likely to introduce a significant
confusion bias as long as its projected separation from the target is more
than 70 mas. Figure~\ref{fig:con_vs_mag_dif} shows the confusion bias as a
function of magnitude difference for a few selected projected separations. 
\section{Applications}
\label{applications}

In this section, we present the results modeling confusion bias for a selection
of fields representative of several of the current set of Key Projects
chosen for SIM subsequent to the first
\textit{Announcement of Opportunity}.\footnote{See the project descriptions
at: \\ \textit{http://planetquest.jpl.nasa.gov/SIM/sim\_team.cfm}.} As mentioned
earlier, the important case of \textit{binaries as signal} will be deferred
to a later paper; here we consider such ``companion'' field stars as
\textit{noise} no matter where they are located in the FOV.

\subsection{Motions of nearby galaxies}

This project (``SIMDOG'') aims at dynamical studies of the Local Group of
galaxies by
measuring their deviations from the Hubble flow.\footnote{The Principle
Investigator for this project is E.J.\ Shaya.} Accurate distances to the
nearby galaxies from e.g.\ cepheids are a part of this work, but another
essential component is the precise measurement of galaxy proper motions. Over
the 5-year lifetime of the mission, repeated SIM observations of the brightest
stars in galaxies out to distances of $\approx 5$ Mpc are expected to yield
these proper motions with sufficient precision.

Initially, it might be assumed that proper motion measurements of a galaxy
with SIM would be insensitive to confusion by extraneous stars in the FOV.
However, if the distribution of brightness over the FOV changes from one visit
to the next, there is a possibility of confusion-induced biases in the proper
motions. One class of changes that has already been evaluated in the work of
\cite{rajbokall01} involves biases caused by slight differences in pointing
the SIM FOV on the target area during successive visits. A second class of
changes in the brightness distribution arises when Galactic foreground stars
with their own proper motions are also present in the SIM FOV; this situation
is likely to be avoidable merely by pointing SIM to a different target star in
the distant galaxy. The third class is not so easily circumvented;
this arises from variations in the brightnesses of stars in the distant galaxy
itself. Variable stars are sufficiently common that one or more of sufficient
relative brightness may occur within the FOV when SIM is pointed at a bright
star in a nearby galaxy, and their variability may not be known ahead of time.
Of course the target star itself may be variable, and this too may be initially
unknown. It is this third class of biases we model here.

\subsubsection{Source model}

We assumed synthetic spectra of galactic supergiants of spectral type A and B
with $T_{eff}$ = 10000~K, $\mbox{log g} = 2$, and $\mbox{log[Fe/H]} = -1.5$.
These are the most favorable stars for SIM observations in the V band, as
described in the key project summary.  

\subsubsection{Results}

The FOV is assumed to consist of the target star and one additional field star,
both of which are stationary.
The position of the target star is measured at two epochs, and the proper
motion bias is estimated as the difference between the two positions. The target
positions will contain some confusion bias owing to the field star and the
granularity of the background galaxy, but if everything remains the same the
calculated proper motion will be zero. The brightnesses of the target and/or the
field star are then changed, and changes in the confusion bias on the target
position measurement will masquerade as a proper motion.

We considered  two cases: (a) the target and a field stars have the same
brightness, but either one or both of them change in magnitude 
by 0.3 mag. between the two visits; and, (b) the field star is fainter than
the target by 1 mag., but one of them changes by 0.3 mag. between visits.
The bias in the proper motion for this case is a function of the magnitude 
difference and the angular separation (projected along the baseline) between
the target and the field star.   

Our simulations indicate that there could be as much as $\approx 800$~\muas\
of bias in the proper motions for galactic supergiants of spectral types A
and B, in case (a) and $\approx 300$~\muas \ in case (b),  if the field star 
is located within a fringe.  If the projected separation of the field star is more than a fringe but less
than 25~mas then the induced bias is  $\approx 30$~\muas \ in case (a) and $\approx 20$~\muas \ in case (b).
If the projected separation of the field star is more than 50 mas, then the bias in
the proper motions owing to brightness variations of 0.3 mag is expected to
be much less than the single measurement precision.
  
\subsubsection{Discussion}

A local group galaxy at a distance of 1 Mpc with a transverse velocity of 100
km/s would have a proper motion of $\approx 20$ \muas/yr, so that biases at the
levels mentioned above could be problematic. Observing strategies to mitigate
such biases will be necessary, such as repeating observations with
changes of a few degrees in baseline orientation and/or observing
several target stars in the same galaxy. As with the other cases of confusion bias which we have modeled here, the impossibility of obtaining a
sufficiently-precise model of the bias precludes any simple correction for it;
the best one can hope for is to be aware of the risks and plan the observations
accordingly.
 
\subsection{Astrometric reference frame tie}

A list of $\approx 1300$ Galactic stars 
spread more-or-less evenly over the sky
will be repeatedly observed with SIM during the mission lifetime. The positions
and proper motions of these ``grid stars'' will enable the definition of an
astrometric reference frame. Unfortunately, this frame may have a small residual
rotation, a possible result of some non-random component to the measured proper
motions of the constituent stars. This ``roll'' component will have to be
removed by measuring the positions and (apparent) proper motions of 50 - 100
distant quasars. These objects are generally too faint to be included in the
regular astrometric grid program of SIM, but they can be measured to sufficient
precision using longer integrations.\footnote{The Principle Investigator for
this project is K.J.\ Johnston.}

The quasars to be used in this study are distributed all over the sky, and are
relatively faint compared to grid stars. These quasars may therefore be
susceptible to confusion from faint (but numerous) A, K, and M stars in the
Galactic foreground, especially near the Galactic Plane. It is this source of
error in the astrometry of faint quasars which will be estimated here. The
effects are exacerbated by the strongly differing SEDs of the quasars compared
to the SEDs of faint red Galactic stars.

\subsubsection{Source model}

The models we adopt for this set of simulations consist of a target quasar,
located for convenience at the center of the SIM FOV, and a field star located
somewhere else within the FOV. Additional structures such as lumpy, extended
jets, are not included. The quasar SEDs are chosen from a small but
representative set of quasar spectra from the ``First Bright Quasar Survey''
\citep{wh00} with redshifts ranging from 0 to 3. These spectra have been
extrapolated wherever they were not adequately
defined over the wavelength range 400 - 1000 nm by assuming a flat spectrum
with values of the nearest known value; they are generally brightest
at the blue end of the spectrum.  The Galactic star can be of
spectral type A1V, K0V, or M6V. We do not include massive O and B type stars,
since these stars are rare at the higher Galactic latitudes where the frame
tie quasars are preferentially found. The foreground star is taken to be 0 - 3
mag fainter than the target quasar, and located at PA = 5\degr, almost directly
``above'' the target. Four different radial separations
between the quasar and the foreground star are modeled (25, 50,
100, and 200 mas), as well as 18 different possible orientations 
(0\degr\ to 170\degr\ in steps of 10\degr\ ) of the
interferometer baseline.

\subsubsection{Results}
\label{qftres}

\begin{figure*}[t]
\epsscale{2.3}
\plottwo{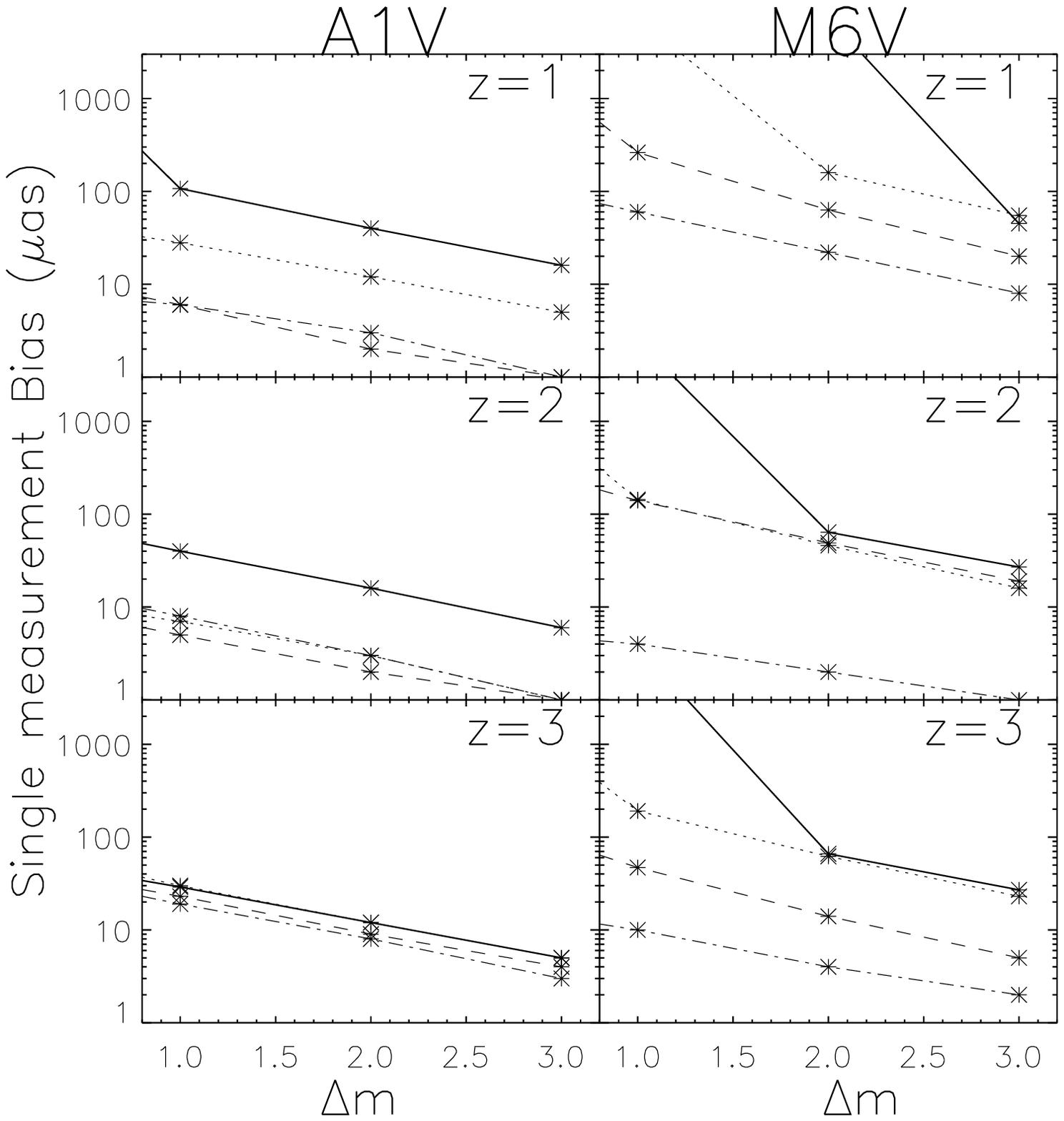}{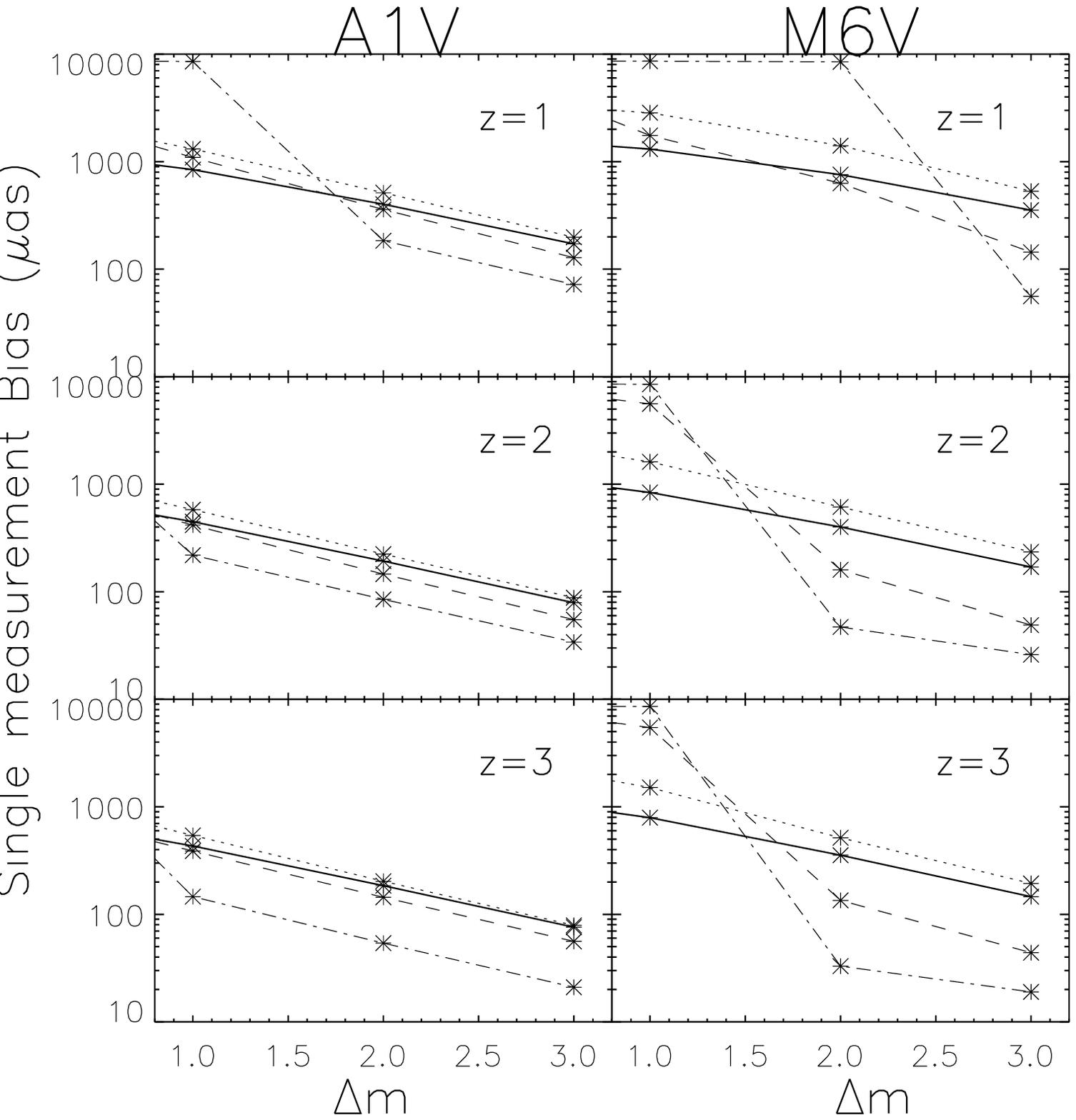}
\caption {Absolute value of bias as a function of $\Delta m$ for field star
separations of 25 (solid line), 50 (dotted line), 100 (dashed line), and
200 mas (dash-dot line) for A1V and M6V field stars. The PA of the field star is 5\degr.
\textbf{Left panel:} Baseline PA = 0\degr. \textbf{Right panel:}
Baseline PA = 90\degr.
\label{fig:qft_plot1}
\label{fig:qft_plot2}
}
\end{figure*}

The left panel of Figure~\ref{fig:qft_plot1} shows the single-measurement confusion bias as a
function of the difference in brightness of the target and the location of
the field star for a baseline position angle of 0\degr.  The right panel of Figure~\ref{fig:qft_plot2} shows a similar
plot for a baseline position angle of 90\degr. Together, these figures show
the typical bounds on the expected biases.

\subsubsection{Discussion}

The simulations show a strong dependence on the SEDs of the quasar and the
field star. For example, M stars cause significantly more bias then A stars
of the same brightness, and there is a (smaller) dependence on the redshift
of the quasar. These biases can be reduced by choosing another baseline orientation as long as $\Delta m \gtrsim 2$ and the projected separation
is $\gtrsim 50$ mas for an A-type field star. The values for an M-type field
star are 3 mag and 100 mas, respectively.

\subsection{Taking the measure of the Milky Way}

SIM will make a major contribution to the study of the distributions of both
dark and luminous matter in the Galaxy through precision measurements of the
distances and proper motions of different classes of Galactic
stars.\footnote{The Principle Investigator for this project is S.R.\ Majewski.}
One of these
studies involves measuring the distances to nearly 100 bright M giants in the
``Baade's Window'' region of the central bulge of the Milky Way. Unfortunately
these are some of the richest star fields ever considered for SIM, and there
are significant concerns about confusion-induced biases in single position
measurements of these bulge stars. We have modeled typical examples of these
fields and investigated whether judicious choices of the orientation of SIM's
baseline can reduce these biases. 

\subsubsection{Source model}
\begin{figure}
\epsscale{1.0}	
\plotone{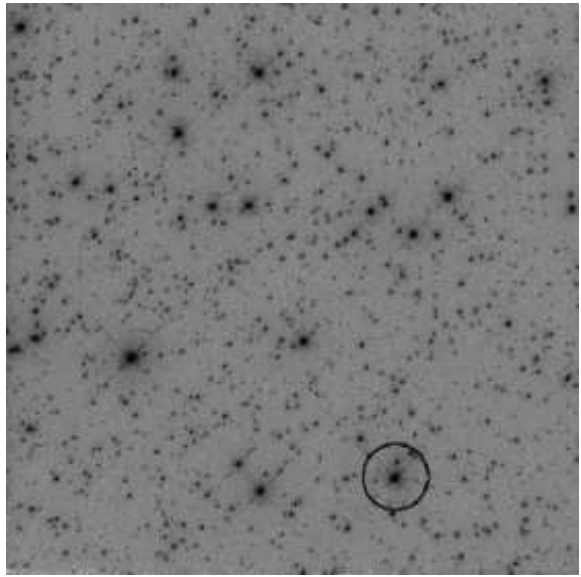}
\caption {
Image of the Milky Way Bulge (Baade's Window) retrieved from 
HST archival data. The image has a scale of 0.0455\arcsec\ per pixel. 
The image has been cropped and displayed as negative.
The black circle indicates the SIM FOV of 3\arcsec. \label{baade}
}
\end{figure}

We have constructed our models for these fields using HST/WFPC2 imaging
observations as a guide. For example, Figure \ref{baade} is an F555W exposure
in the region of Baade's window (HST Proposal ID 8574; Target name
FIELD180310-295143). This image has a scale of 0.0455\arcsec\ per pixel.
We obtained photometric and astrometric information for more than 3800 stars
in this image from J.\ Holtzman (private communication); his catalog included
V \& I magnitudes along with associated errors, X \& Y position measurements,
and the RA \& DEC (J2000) of the stars. We abstracted a set of $\approx 40$
target FOVs from this image, each centered on a bright ``target'' star, and
constructed an idealized model for each field consisting of a list of all
stars lying within and just outside of the SIM FOV.
The SED of each star was estimated as follows:
First, we fitted a 2nd degree polynomial to a theoretical plot of V-I color
index vs.\ effective temperature obtained from Kurucz's 
models\footnote{{\it http://kurucz.harvard.edu/grids/gridP00ODFNEW\linebreak[1]/rijklp00k0odfnew.dat}}
assuming that all the stars are solar-type dwarfs.
This yielded a relation between the effective surface temperature and the V-I color
index of the star. Using the known V-I colors for all stars in the catalog,
we estimated the corresponding surface temperatures and hence the spectral
types. A large fraction ($\approx 85\% $) of the V-I colors lie in the
theoretically-expected range of 0.5 - 2.57; this is consistent with (but does
not prove) the assumption that these stars have essentially zero metallicity.
We also ignored any extinction corrections. All stars lying within a circle
of 6\arcsec\ of the bright target star were considered as potential sources of
confusion, since some light from these stars can be diffracted into the FOV.
A baseline orientation was then chosen in the range 0\degr\ - 170\degr\ in
steps of 10\degr, and the position of the target star calculated first without,
and then including, the surrounding stars.
The difference is the confusion-induced single-measurement astrometric bias. 

\subsubsection{Results}

\begin{figure}
\epsscale{1.0}	
\plotone{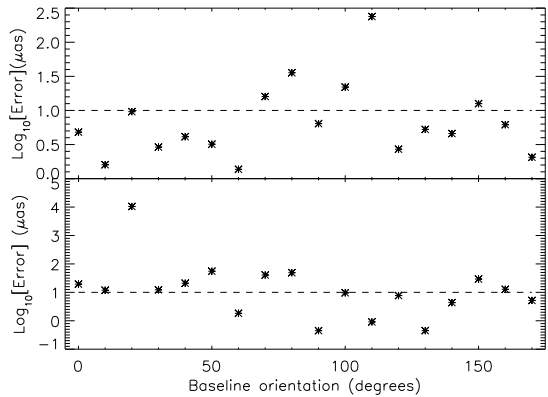}
\caption{Single-measurement confusion-induced astrometric bias as a function
of baseline orientation for models of two very crowded fields in
Baade's Window. Note the logarithmic scale on the $y$ axis; the biases can
be large for some baseline orientations. The dashed horizontal line is drawn
at a bias of 10 \muas, which is roughly the estimated single-measurement error
expected from noise and instrumental instability. \label{bovserr_baade}
}
\end{figure}

Of the 40 target fields we have simulated, 22 show biases in excess of
10 \muas\ if the same baseline orientation is chosen (90\degr\ for these
simulations). The bias can be reduced below 4 \muas\ in 39 out of 40 targets
simply by choosing a suitable baseline orientation. This suggests that some
experimentation with the actual data set could therefore be useful, by taking
the data at several baseline position angles and rejecting anomalous points.
Figure~\ref{bovserr_baade} shows an example of the astrometric bias as a
function of baseline orientation for two different fields. For the first field
(top panel), 12 out of 18 baseline orientations show biases less than
10 \muas. For the  second field (bottom panel), 7 out of 18 baseline orientations show correspondingly small biases. Also, note that the biases
can occasionally be very large (e.g., at baseline PA = 20\degr\ in the
second field).

\subsubsection{Discussion}

Our results generally show that, in these very crowded fields, the amount of
bias grows with the number of field stars. This is expected, as consideration
of the phasor diagram in Figure \ref{fig:phasor_diagram} will show. Furthermore,
the largest contributions to the net bias are primarily caused by a small number
of judiciously-situated field stars. These stars can even be located outside
the FOV.
\begin{figure*}
\includegraphics[width=5.5cm]{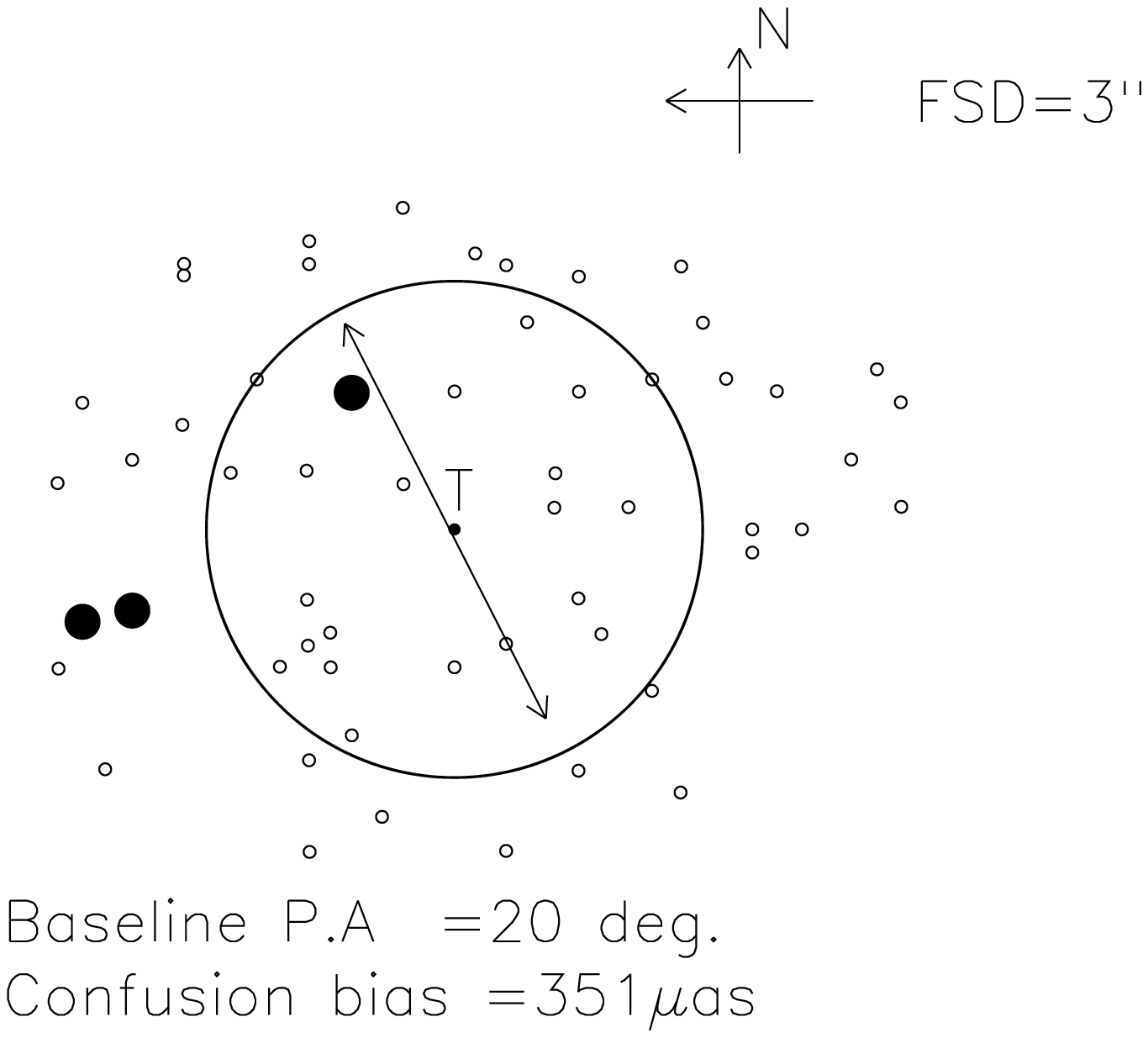}
\includegraphics[width=5.5cm]{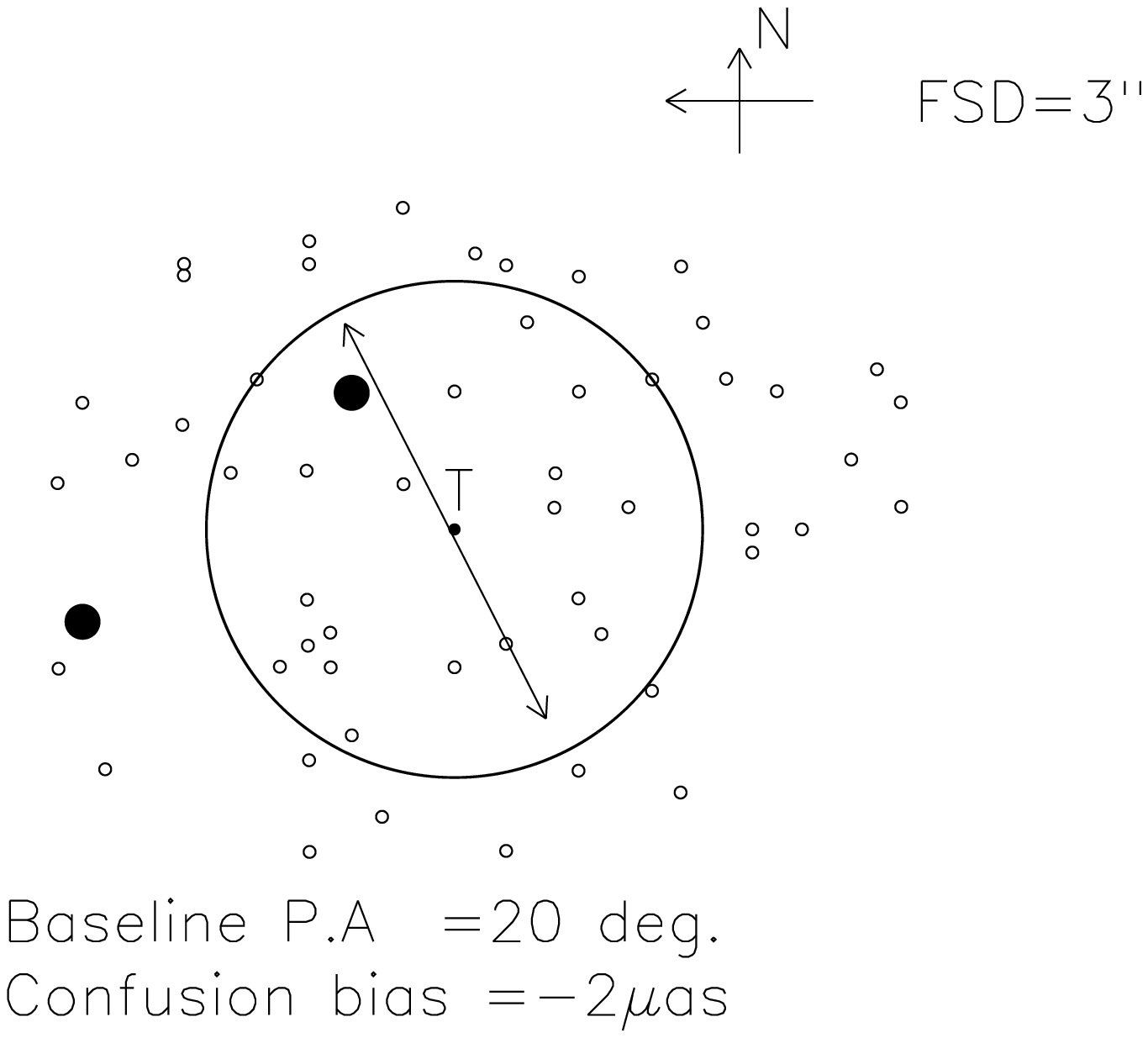}
\includegraphics[width=5.5cm]{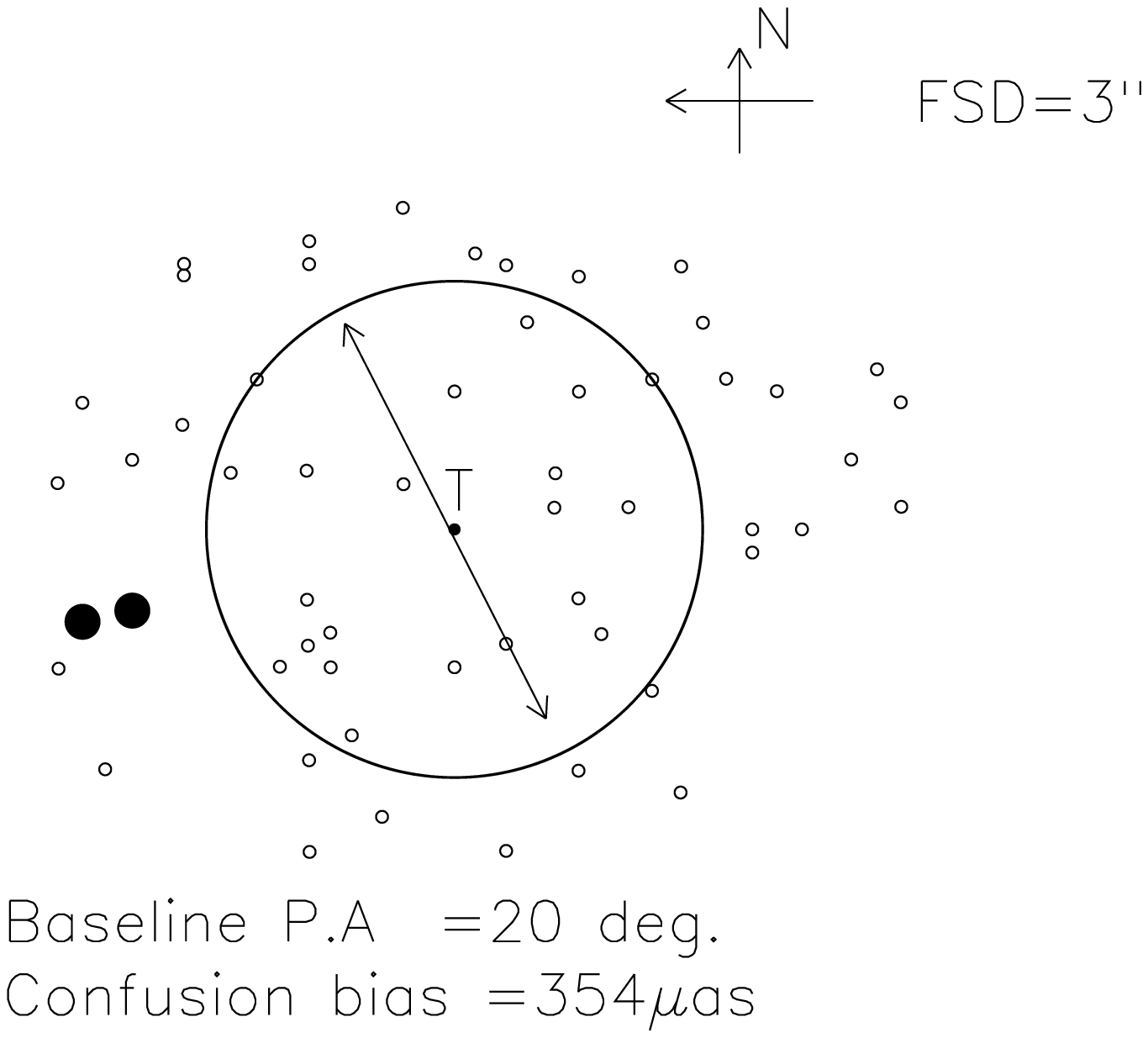}
\caption {\textbf{Left panel:}
One of the two field stars at position angle of 110\degr\ outside
the field-stop, shown by filled dots, has a projected distance of about 2 mas
and is fainter by 1.4 magnitudes.
\textbf{Middle panel:} Excluding it from the model reduces the bias from
351 \muas\ to -2\muas. The other field star has a projected distance of about
13 mas, but it is fainter by 6.8 magnitudes; excluding it from the model makes
little difference. \textbf{Right panel:} Somewhat surprisingly, a field
star located well inside the FOV at a projected distance of 1.3\arcsec\ and
brighter than the target by 0.95 mag also makes a negligible contribution to
the net bias.
\label{model_34}}
\end{figure*}
Figure~\ref{model_34} is an example of such a case. In this field, there are
56 field stars in our model. Two of them have projected distances of only 2 mas
and 13 mas, and are fainter by 1.4 and 6.8 mag respectively, but they are both
located outside the edge of the FOV and, at first sight, should not be a
problem. Nevertheless, the simulations show that the measured target position is
biased by 351 \muas. When the field star at 2 mas is excluded, the bias drops
from 351 \muas\ to -2 \muas. Further excluding the 13 mas field star makes very little difference; this star is probably too faint to cause problems, or it is
at a fortuitous location on the fringe pattern. There is a third field star
which would seem a likely candidate for causing a bias; it is \textit{brighter}
by 0.95 mag and located \textit{inside} the FOV, at a projected distance of
1.3\arcsec. However, excluding this star makes little difference to the bias,
as is shown in the right panel of Figure \ref{model_34}.

\subsection{Cluster distances}

This project is aimed at measuring the distances to a set of selected open and
globular clusters in the Galaxy in order to determine their precise ages by
combining SIM parallax data with other existing data.\footnote{The Principle
Investigator for this project is G. Worthey.} A few stars from each of
these open and globular clusters will be chosen as targets for SIM astrometry.
However, these targets are necessarily in crowded stellar fields, and the
examples of the previous section suggest that the confusion-induced biases
could be large. We have modeled these biases
in the typical case of NGC~6440, an old globular cluster at a distance of
$\approx 8.4$ kpc, and located near the Galactic center at l = 7.7\degr,
b = +3.8\degr. NGC~6440 is a well-studied globular cluster, with a core radius
of 8\arcsec. It is one of the 14 globular clusters known to have X-ray emission. 
Specifically, we have addressed the following questions with our models:
\begin{itemize}
\item What is the typical confusion bias for the 50 brightest targets in
NGC~6440?
\item At what area density of bright field stars are the target biases
tolerable?
\end{itemize}

\subsubsection{Source model}

\begin{figure}[ht]
\epsscale{1.0}
\plotone{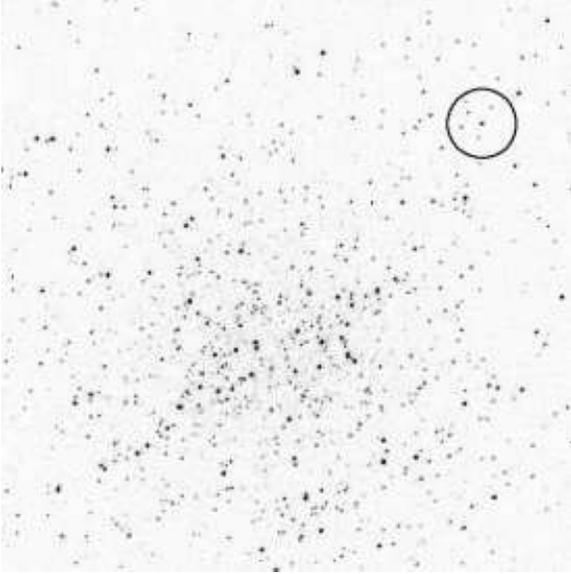}
\caption{PC image of the Globular Cluster NGC~6440 cropped and displayed as a
negative. The image scale is 0.046 \arcsec/pixel. The black circle indicates
the SIM FOV of 3\arcsec.  \label{globular}
}
\end{figure}

\begin{figure}[ht]
\epsscale{1.0}
\plotone{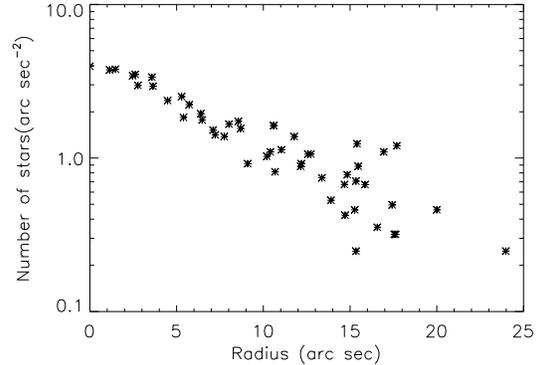}
\caption {Number density of stars as a function of radial distance from
the core of the cluster NGC~6440.  The magnitudes of the stars lie in the
range from 16.19 to 25.52 \label{ngcres1}
}
\end{figure}

Our model for NGC~6440 is based on an HST image obtained with the Planetary
Camera of WFPC2, as shown in Figure \ref{globular} and kindly provided by
G.\ Worthey.
%
Most of the bright stars in this picture are K giants; we generated a catalog
of these stars using the `SExtractor' software.\footnote{Bertin, E.,\\ 
{\it http://terapix.iap.fr/rubrique.php?id$\_$rubrique=91}} Although it is well
known that this approach does not provide accurate photometry in crowded fields,
our goal here is not to obtain precise corrections to the astrometric
observations (which would require micro arc-second positions as well as accurate
magnitudes) but rather to elucidate the general level and nature of the biases
which may be expected. The number of sources which can be obtained with this
software depends on the assumed value of the background intensity; after
a few trials, we settled on a background value of $1.5 \times \sigma_{rms}$,
which yielded about 2400 stars with positions and V magnitudes. A visual
comparison of the selected objects with the original image showed a good
correspondence. The area density of these stars as a function of the radial
distance from the center of the cluster is shown in Figure \ref{ngcres1}.
We then selected 51 circular fields at different locations
in the cluster (one is shown in Figure \ref{globular}); each field is centered
on a bright star (presumed to be a K-giant). The model constructed for that
field consists of a list of the positions and magnitudes of the target star
and all field stars within a radius of 3\arcsec\ of the target; this allows
for diffraction of light into the 3\arcsec\ (diameter) SIM FOV from stars which
are located beyond the edge of the aperture.

The SEDs of the model stars were determined from estimates of surface
temperature, gravity, and metallicity, as follows. From the V magnitudes
(with the assumption that all stars are K giants) we attributed a temperature
to a star randomly in the range 4000-4750 K (4000, 4250, 4500,4750) and a
surface gravity randomly in the range of 1.5 to 4 with a step size of 0.5.
We assumed that all the stars have solar metallicity. The spectra so determined
were renormalized to account for the distance to the cluster and the magnitude
of the field star.

\subsubsection{Results}

\begin{figure}[h]
\epsscale{1.0}
\plotone{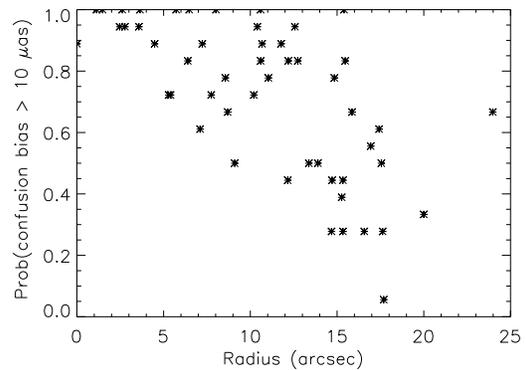} 
\caption {Probability that the  confusion bias exceeds 10 \muas \ in a single measurement as a
function of radius of the globular cluster NGC6440.  
\label{ngcres4}
}
\end{figure}
\begin{figure}[ht]
\epsscale{1.0}
\plotone{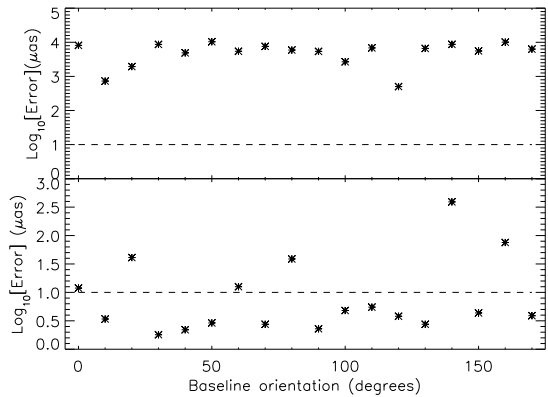}
\caption {Confusion bias as a function of baseline orientation for model fields
located at 3.6\arcsec\ (top panel) and 20\arcsec\ (bottom panel) from the
center of NGC~6440. Target fields at the periphery of clusters are clearly
preferable, but may still have an appreciable probability (here $\approx 30\%$)
for a substantial confusion bias. \label{bovserr_ogc}
}
\end{figure}

In each of the 51 fields we obtained the confusion bias for 18 different
orientations (position angles) of the SIM baseline (0\degr\ to 170\degr\ in
steps of 10\degr).  
The fields used in this set of simulations typically have target stars at V $\approx$ 16-18, with brightest field stars at V $\approx 16$ or fainter, 
down to $\approx$ 25 (with some exceptional cases; 
in 5 out of 51 fields, brightest field stars were 
brighter than the target by 1.2, 1.6, 3.2, 2, 4.6 magnitudes; 3 of
them were within the core radius and the remaining two were within 16 arcsec
radius.). 
Of all the cases we analyzed in each field, 20\% of them showed biases in excess of
10~\muas \ in all 18 baseline orientations considered; 
Figure~\ref{ngcres4} shows the probability of a single
measurement confusion bias exceeding 10~\muas \ as a function of
radius of the cluster. For a given field, if the confusion bias was more than 10~\muas \ in
12 out of 18 orientations, the probability was estimated as  2/3 at the radial distance of
the field from the center of the globular cluster.
This figure indicates that, for NGC6440, the probability of having large
confusion bias can be as large as 95\% even at the radius of 15\arcsec \ and there 
could be as much as 40\% probability of large confusion bias at large radial distances.
The probability
of obtaining large confusion bias drops at larger radial distances in the cluster, as
the local stellar surface density decreases. This latter point is illustrated
in Figure~\ref{bovserr_ogc}, which shows the confusion bias as a function of
baseline orientation for two different fields, located at radial distances of 
3.6\arcsec\ and 20\arcsec\ from cluster center. For the field at 3.6\arcsec\
(which is actually inside the cluster core), the biases are always large (note
the logarithmic scale) at all baseline orientations. For the field at 20\arcsec,
the error is larger than 10 \muas\ in 30\% of the baseline orientations.

\subsubsection{Discussion}

The statistics of these results indicate that the probability of having an
unacceptable level ($\ge$~10~\muas) of confusion bias in the \textit{positions} of target
stars is will exceed 30\% for a single
observation carried out at any random baseline orientation if the stellar
density in the area exceeds 0.4 stars per square arc-second (compare Figures~\ref{ngcres1} and ~\ref{ngcres4};
In other words, about 3 stars within the 3\arcsec \ FOV is acceptable). 

However, the focus of this particular Key Project is on \textit{distances}
to objects, and hence on the parallaxes of the target stars. How do our results
apply to the measurement of parallaxes? Such a measurement could be carried out
with SIM in a special way, namely, ensuring that the baseline orientation was
identical for observations taken $\approx 6$ months apart in time.
This is perhaps
possible in principle, but very unlikely in practice. First, the baseline
orientation can not easily be set to $\approx 1$\degr; indeed,
it is in fact generally not even known until the baseline
vector has been calibrated. This
means that parallaxes will suffer from the same magnitude of confusion bias
as do regular position
measurements. In the case of the cluster Key Project, a possible strategy
could be to choose a number of target stars each at some large distance from
the cluster center (such that the stellar density is lower than the limit cited
above), and obtain parallaxes on each of them using the standard SIM observing
strategy.\footnote{This strategy is still under development and includes a complex calibration program.} Discordant single-measurement data points would
then simply be rejected from the data set as a likely consequence of confusion.

\section{Limitations of the models}
\label{limitations}

The first, and perhaps most painful, limitation of the modeling we have
described here is that, in spite of our apparent ability to compute the
confusion bias in any single measurement with high precision, the results
are in general not likely to be sufficiently accurate to provide actual
corrections to the target positions. The simple reason for this is that
the \textit{true} field star positions are not known with accuracies of
the same order as the expected single-measurement precision of SIM,
$\approx 10$ \muas. Whether the modeling might be sufficient in any
particular case will depend on the specific details of that field on the
sky; our modeling tools (described in the next section) can then be used
to evaluate the biases by e.g.\ varying the positions and brightnesses of
the field stars over plausible ranges.

The second limitation of our modeling concerns the instrument model we have
adopted for SIM. There are several simplifications we have made which
could be problematic:
\begin{enumerate}
\item
In our numerical simulations, we have assumed that the FOV is exactly
circular, and that all the photons diffracted into it will be collected
(with some efficiency) by the detector. However, in practice the focal
plane of SIM's camera will be covered by a CCD detector with pixels of
$\approx 2$\arcsec\ on a side, and it is presently planned to average the data in
three neighboring rows.
\item We have assumed that the photons collected in each FOV will be dispersed
in the camera in some way into 80 channels with central wavelengths and bandwidths as specified in Appendix \ref{appsec:params}. In fact, a thin prism
will be inserted into the light path before imaging onto the CCD detector,
turning the camera into an objective prism spectrograph. Such instruments
require special calibration and are subject to a degree of internal confusion
caused by overlap of spectra from field stars with the spectrum of the target
star.
\item  We have estimated the fringe parameters (total power, fringe amplitude
and fringe phase) analytically, and neglected the details of just how they will
be measured on-board the spacecraft.
\item We have assumed that the throughput of the system is the same for
different locations in the FOV. In fact, the presence of the prism will cause
the throughput (and the dispersion) to change depending upon the angle of
incidence at the prism, and hence to be different for the target and for the
field stars.
\end{enumerate}

We have carried out a further study using a more sophisticated instrument model
which removes the first three of these limitations. The details of this extended
model will be described elsewhere \citep{sriron07}, but we give here the results
of repeating the bias computations on a subset of the fields we have described
in the present paper. Table~\ref{comp-table} compares the biases obtained using
the ``simplified'' approach presented in this paper with those obtained using
the more detailed model of the focal plane. The cases listed are taken from the
Quasar Frame Tie key project (cases 1 - 5) and a selection of binary models
(cases 6 - 8). Perhaps not surprisingly, the differences are
roughly proportional to the magnitude of the computed bias, although these
differences are generally at a level of only a few percent. Unfortunately, the
utility of the results from this more sophisticated instrument model is still
compromised by inaccurate input data on the field stars, and since it also comes
with considerable additional complexity we have not used it further in this
paper.
\begin{deluxetable}{ccccc}
\tablecaption{Comparison of the single-measurement confusion bias obtained
with the present simplified model and with the more detailed model of
\citet{sriron07}.\label{comp-table}}
\tablewidth{0pt}
\tablehead{
\colhead{Case} & \colhead{Model} & \colhead{Simple} & \colhead{Detailed} &
\colhead{Difference} \\
\colhead{Number} & \colhead{parameters} & \colhead{(\muas)} & \colhead{(\muas)}&
\colhead{(\muas)}
 }
\startdata
1    & Q(z=0), A1V, 2, 50, 5, 90   &\phn442 &\phn431 &11\\
2    & Q(z=0), M6V, 2, 50, 5, 90   &1072 &1036  &36\\
3    & Q(z=2), A1V, 2, 50, 5, 90  &\phn224      &\phn215     &\phn9\\
4    & Q(z=2), M6V, 2, 50, 5, 90  &\phn614     &\phn592     &22\\
5    & Q(z=2), M6V, 2, 50, 5, 10 &\phn$-46$    &\phn$-46$    &\phn0\\
6    & A1V, B1V, 3, 1500, 10, 90  &\phn\phn$-1$     &\phn\phn\phn0    &\phn1\\
7    & A1V, M6V, 2, 50, 90, 90  &\phn$-59$     &\phn$-61$      &\phn2\\
8    & A1V, M6V, 2, 25, 90, 90  &\phn$-53$     &\phn$-46$     &\phn7\\
\enddata
\tablecomments{In column 2, Q(z=0), A1V, 2, 50, 5, 90 means target quasar
with redshift 2, A1V field star, $\Delta$m = 2, radial distance 50 mas,
PA = 5\degr, baseline orientation 90\degr.}
\end{deluxetable}

\section{Modeling Tools}

The source and instrument models described here have been implemented in a suite
of programs written in the IDL programming system.
These programs are available\footnote{{\it http://www.stsci.edu/$\sim$rjallen/sim/}}
for further experimentation, including the code developed for
estimating confusion bias in the specific target fields described in this
paper.

\section{Summary and Discussions}
\label{sum}

We have examined the bias that can occur in a single measurement with SIM
owing to the presence of field stars within the FOV. In order to accomplish
this task we have presented a model for the SIM interferometer, and
a description of how SIM carries out a single measurement of the position of an isolated target star.
The measured instrument response is then perturbed by adding a field star
to the model FOV; the difference in the angles measured in the two cases is called the ``confusion bias''. The extremes of this bias are calculated for
the specific (but common) case of a binary system in order to illustrate its
main properties.

A number of source models are then developed which resemble the fields to be
studied by SIM in several of the Key Projects already selected for inclusion
in the initial mission science program. An unconfused version of the source model consisting only of the main target star is used as a reference
measurement, and the results compared with a measurement made on the
fully-populated field. The difference is the confusion bias in a single SIM
measurement. Observations are simulated at various orientations of the
interferometer baseline, and variants of the full field are examined in order
to understand the sensitivity of the bias to structural details in the field.

The magnitude of the confusion bias is found to depend on a number of factors,
some obvious, others perhaps less so:
\begin{itemize}
\item the relative brightnesses of the target and the field stars;
\item the shapes of the SEDs of the target and field stars; 
\item the angular separation of the stars from the center of the FOV;
\item the angular separation of the field stars from the target star as
projected on the interferometer baseline; and,
\item the baseline orientation.
\end{itemize}
The largest contributions to the confusion bias in a crowded field come from
a small number of stars having small projected angular separations from the
target, but these stars may actually be located outside of the FOV. Field stars
which are less than 4 mag fainter than the target and which have projected separations within 100 mas of the target are potentially the most troublesome.

The results of this study provides the understanding and the tools required
to examine the likelihood of confusion bias in any single measurement with SIM.
Unfortunately, data on the field stars in any specific FOV (especially their
positions) is not likely to be available with sufficient accuracy to actually
remove this bias.\footnote{There are some possible exceptions one could
imagine, but we have not explored them further here.} Our study nevertheless
suggests some strategies for recognizing the presence of confusion bias and for
dealing with it, both in the observation planning stage and in the data
reduction stage. These strategies might include the following:
\begin{itemize}
\item While dealing with crowded fields, avoid fields with star densities in 
excess of 0.4 stars per square arcsec. 
\item If avoidance is impossible, evaluate the likelihood of confusion in the
field by using the tools developed here.
\item If confusion is likely, try to reduce your sensitivity to it by planning
the observing program so that data is taken at the least sensitive orientations
of the interferometer baseline.
\item If too little is known about the specific field, plan to distribute the
available observing time over a number of orientations of the interferometer
baseline which differ by a few degrees from each other. Inconsistent values
in the data set can then be rejected with motivation.
\item If confusion is suspected in a given set of observations for which
no prior data exists, acquiring new imagery from e.g., speckle or adaptive optics
imaging would be useful for building a model.
\end{itemize}
There is one additional strategy suggested by our more accurate model of the
SIM focal plane. The CCD detector in the focal plane of
SIM's camera is planned to have pixels which are smaller than the diameter
of the FOV. If the data in the individual pixels can be made available, it would
be possible to choose e.g.\ only the central pixel, thereby effectively
reducing the FOV and possibly attenuating an offending field star. The penalty
of fewer target photons could then be offset by a reduction in the level of
confusion bias. This possibility will be discussed in more detail in a future
paper \citep{sriron07}.

We wish to emphasize that the results of this paper refer to a bias which may
be present in a single measurement of angular position with SIM. The
determination of the full set of astrometric parameters (position, parallax,
proper motion) on any SIM target will be done with a number of
measurements, reducing the effects of any single anomalous point. Furthermore,
the ultimate accuracy of the results depends on an extensive calibration program
to determine the instrumental parameters, including the baseline length and
orientations for each field observed. 

\acknowledgments

We are grateful to our colleagues in the SIM Science and Engineering Teams
for discussions about SIM and about the potential for confusion-induced bias
in SIM measurements. We thank Xiao Pei Pan, Mike Shao, and Jeff Oseas of JPL
for providing the
latest values of the instrumental parameters for SIM. Information on the
different Key Project fields modeled here was provided by Carl Grillmair,
Guy Worthey, Ed Shaya, Rick White, Jon Holtzman, and Norbert Zacharias.
This work was funded primarily by the SIM project office at JPL under
contract \#1268384 and carried out at the Space Telescope Science Institute;
the paper was written with the partial support of the STScI Director's
Discretionary Research Fund. Finally, our sincere thanks to an anonymous
referee whose critical comments helped to improve the presentation.


\appendix

\section{Michelson interferometer response}
\label{appsec:response}

Michelson interferometers are used in astronomy at wavelengths from the radio
to the optical for spectroscopy, for astrometry, and for synthetic imaging.
The major equations giving the response of such interferometers for the latter
two applications are summarized here.

\subsection{Interferometers for astrometry and imaging}

At any given wavelength $\lambda$ and baseline separation $B_{ij}$
(both e.g.\ in meters), the response of a simple adding Michelson
interferometer of the type used for astrometry and imaging of partially
coherent light in optical astronomy can be modeled as
(e.g.\ \citet{bw75}, p.\ 491 \textit{et. seq.}):
\begin{equation} \label{eqn:generalresponse}
P(\Delta_{ij}, k) = P_{0} [ 1 + A \cos(2 \pi k \Delta_{ij})],
\end{equation}
\noindent where $P_0$ is proportional to the target brightness
(e.g.\ in photons/sec), $A \leq 1 $ is called the \textit{fringe amplitude}
and depends on the target structure, $k = 1/\lambda$ is the \textit{wavenumber}
(in e.g. meters$^{-1}$), and $\Delta_{ij}$ is the total
\textit{optical path difference OPD}
(e.g.\ in meters) between the two sides (or ``arms'') of the
interferometer.  For point source targets, $A = 1$, and the normalized version
of equation \ref{eqn:generalresponse} can be considered as the interferometer's
far-field ``point spread function'' (PSF), similar in concept to the PSF of a
filled-aperture telescope.

\begin{figure}
\epsscale{0.8}
\plotone{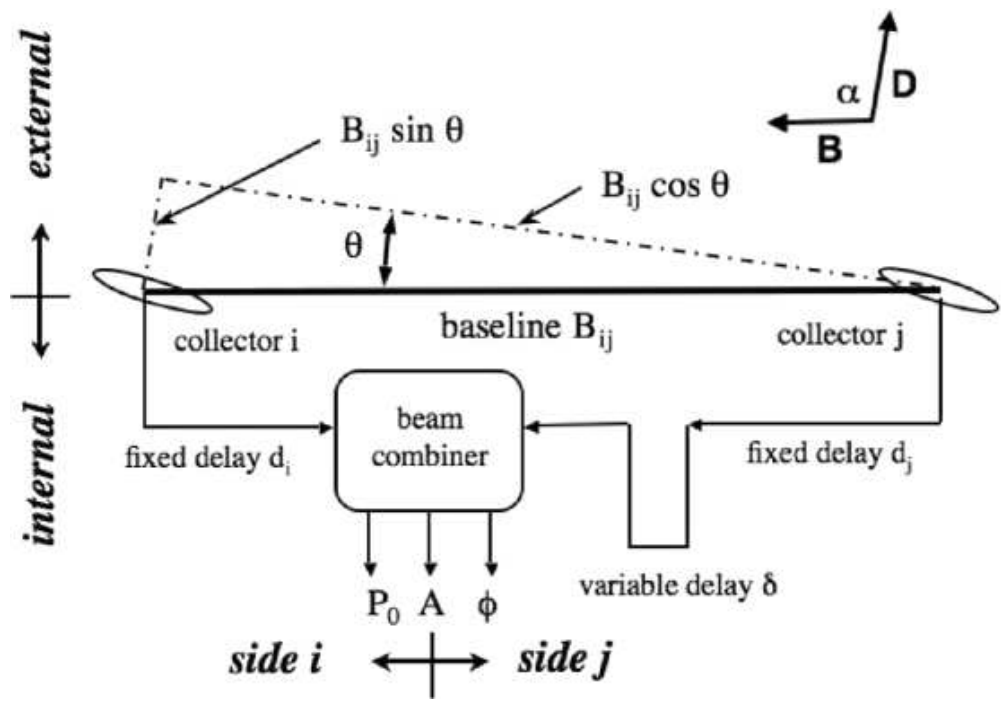}
\caption{Sketch of the basic Michelson interferometer as used for astrometry
and imaging in radio and optical astronomy. \textbf{\textsf{B}} is a vector
in the direction of the baseline, and \textbf{\textsf{D}} a unit vector in
the direction in which we want to compute the interferometer's response.
$\theta = \alpha - \pi / 2$, and P$_0$, A, and $\phi$ are defined in the text.
\label{fig:generalsketch}}
\end{figure}

The total OPD $\Delta_{ij}$ consists of components on each side $i$ and $j$
of the interferometer, and each of those consists of an \textit{internal} and
an \textit{external} part. Furthermore, some of the internal delay is
``fixed'', and some may be ``variable''. Referring to Figure \ref{fig:generalsketch}
we can write:
\begin{eqnarray} \label{eqn:delaycomps}
\Delta_{ij} & = & [\mbox{total side $j$ delay}] - 
                  [\mbox{total side $i$ delay}], \nonumber
\end{eqnarray}
where the components on each side are:

\begin{center}
\( \begin{array} {rcl}
\mbox{side $j$ internal delay} & = & \mbox{fixed delay + variable delay} \\
  & = & d_j + \delta , \mbox{ and} \\
\mbox{side $j$ external delay} & = & 0 \mbox{ by construction. Further,} \\ 
\mbox{side $i$ internal delay} & = & \mbox{fixed delay + variable delay} \\
  & = & d_i  +  0 \mbox{ by construction, and} \\ 
\mbox{side $i$ external delay} & = & B_{ij} \sin \theta .
\end{array} \)
\end{center}

\noindent This last equation is simple geometry, as Figure \ref
{fig:generalsketch} shows, but it is often stated as the dot product of a vector
\textsf{\textbf{B}} parallel to the direction of the baseline with a (unit)
vector \textsf{\textbf{D}} in the direction for which we wish to compute $\Delta_{ij}$. \textsf{\textbf{D}} makes an angle of $\alpha$ with
\textsf{\textbf{B}}, see Figure \ref{fig:generalsketch}.
Then the ``side $i$ external delay'' can be written as
\textsf{\textbf{B}} $\bullet$ \textsf{\textbf{D}} $ = B_{ij} \cos \alpha $.
Defining $\alpha = \theta + \pi / 2$ we have ``side $i$ external delay''
$ = B_{ij} \sin \theta $, with $\theta$ defined as shown in Figure \ref{fig:generalsketch}; in fact, $\theta$ is the half-opening angle
of a cone with axis parallel to the interferometer baseline. Note that $\theta$
is also the angular distance of the direction of interest from a direction
perpendicular to the baseline ($ - \pi / 2 \leq \theta \leq + \pi / 2$).
Our idealized
interferometer has a constant response in directions orthogonal to the
baseline; the actual field of view will be further restricted by the
practicalities of its design. Note further that equation
\ref{eqn:generalresponse} also closely describes the
pattern in \textit{transmission}, such as would occur if a laser (or CW radio
transmitter) would be sent from an appropriate point in the beam path
\textit{backwards} through the interferometer. This is an
expression of a general \textit{reciprocity theorem}, familiar in radio
engineering, which is itself rooted in the symmetry of the wave equation
for the electromagnetic field with respect to the direction of time.

Finally, equation \ref{eqn:generalresponse} applies only at a single wavelength;
one must also account for the finite bandpass. In essence the response becomes
a sum of many different patterns within the bandpass, and the coefficient of the
``cos'' term in equation \ref{eqn:generalresponse} will be multiplied by
a factor which is essentially the Fourier Transform of the band shape for that
specific channel. Without detailed knowledge of that band shape we can not
calculate the final form of equation \ref{eqn:generalresponse}; however, the
results are likely to be amenable to a simple parameterization, using the
following definitions. The mean wavenumber $\overline{k_n}$
for channel $n$ is defined as:
\begin{equation} \label{eqn:MeanWaveNumber}
\overline{k_n} = \frac{\int k F_n(k) d k}
     {\int F_n(k) d k} ,
\end{equation}
where $F_n(k)$ is the band shape of channel $n$ expressed in wave numbers
($k = 1/\lambda = \nu / c$, where $\nu$ is the frequency in Hz).
The \textit{coherence length}
$\Lambda_n$ (e.g.\ in meters) for channel $n$ is defined here as:
\begin{equation} \label{eqn:correlationlength}
\Lambda_n \approx 1/(\Delta k)_n
\end{equation}
where $(\Delta k)_n$ is the full width at half maximum of an assumed
Gaussian band shape $F_n(k)$ centered at $\overline{k_n}$. The coherence length
is the scale size of the wave packet formed by the collection of photons at
various neighboring wave numbers that make up the broad-band signal. It is
given by the expression $\Lambda_n = \overline\lambda^2/\overline{\Delta\lambda}$.
With these definitions we can write our model expression
for the fringe pattern in channel $n$ as:
\begin{equation} \label{eqn:almostfinalresponse}
P(\Delta_{ij}, \overline{k_n}, \Lambda_n) = P_{0} \{ 1 + 
          A e^{ -C \left( \frac{\Delta_{ij}}{\Lambda_n} \right) ^2 }
          \cos ( 2 \pi \overline{k_n} \Delta_{ij} ) \}
\end{equation}
where $C = \pi ^2 / 4 \ln 2 = 3.56$.
Suppose further that we design the interferometer such that $d_i = d_j$
(or define $\delta = 0$ such that this occurs), then the OPD becomes:
\begin{equation} \label{eqn:finaldelay}
\Delta_{ij} = \delta - B_{ij}\sin\theta .
\end{equation}
Inserting this last relation into equation \ref{eqn:almostfinalresponse}
gives the final (approximate) expression for the interferometer response
in a single channel of finite bandwidth:
\begin{equation} \label{eqn:finalresponse}
P(\delta, \theta, \overline{k_n}) = P_{0} \{ 1 + 
  A e^{ -3.56 \left[\frac{\delta - B_{ij}\sin\theta}{\Lambda_n} \right] ^2 }
  \cos ( 2 \pi \overline{k_n} [ \delta - B_{ij}\sin\theta ] ) \}.
\end{equation}
Considered as a function of $\theta$ this equation describes the fringe pattern
anywhere in the sky for a given value of internal delay $\delta$. Alternatively, equation \ref{eqn:finalresponse} describes the response to a source located at a
fixed $\theta$ in the sky as a function of the internal delay $\delta$. 

Equation \ref{eqn:finalresponse} needs to be modified in order to render
it specific to SIM. First, SIM's optical beam combiner effectively adds an
additional $\pi / 2$
of phase delay to one of the incoming beams, turning the ``cosine'' function
into a ``sine''. Second, SIM always ``observes'' in a direction closely
perpendicular to the baseline orientation (cf.\ Figure \ref{fig:generalsketch}),
so that $\theta$ is small and we can set $\sin \theta \approx \theta$. Third,
the exponential term describing the amplitude decorrelation can be set to
unity, as the bandwidths of the individual channels are narrow.
With these changes, and ignoring the subscripts $ij$ of the baseline $B$,
we can write equation \ref{eqn:finalresponse} for any channel with mean
wavenumber $\overline{k}$ as:

\begin{equation} \label{eqn:finalresponse1}
P(\delta, \theta, \overline{k}) = P_{0} \{ 1 +
  A \sin ( 2 \pi \overline{k} [ \delta - B\theta ] ) \}.
\end{equation}

\section{Instrument and Model Details}
\subsection{Current SIM parameters}
\label{appsec:params}

Our SIM instrument model assumes the following values for the various
parameters required in the simulation of the interferometer response:
\begin{description}
\item{\it Baseline length:} 9.000 m for the astrometric interferometer. 
\item{\it Collector size: } Siderostats, outer diameter 304.5~mm, inner diameter 178~mm, net area = 479.378 cm$^2$.
\item{\it Field stop size:} Nominally 3\arcsec~diameter. However, this can be
chosen in the simulation code to be any one of four possible values
(1,2,3 or 4\arcsec).
\item{\it Number of channels and bandwidth:} The design for the fringe
disperser in SIM has 80 narrow-band channels with bandwidths progressively
increasing from 1.8~nm at 401.9~nm, to 24.9~nm at 985.5~nm. The band shape
is likely to be approximately Gaussian, and this is what we have used;
however, the simulation code allows the user to choose a rectangular
passband instead. The list of central wavelengths and bandwidths assumed
for these simulations can be found at the author's web site.\footnote
{{\it http://www.stsci.edu/$\sim$rjallen/sim/}}
\item{\it Throughput ($Th$):} The Throughput varies across the 80 channels,
depending on the reflection/transmission optics and the QE and spectral
response of the detector.  We found that the experimentally-measured values
could be modeled best using the relation $Th = \sum_{i=0}^7 c_i \lambda^i$
with $c_0 = 18.2$, $c_1 = -338.5$, $c_2 = 2076.2$, $c_3 = -6239.4$,
$c_4 = 10463.3$, $c_5 = -10029.1$, $c_6 = 5151.4$ and $c_7 = -1102.1$, for
$\lambda$ expressed in microns.

\item{\it Pointing accuracy:} There are two parts to the ``pointing''
of SIM: The \textit{angle tracker} will center the target in the instrument
FOV and superpose the images from each side of the interferometer to within
10~mas for bright targets $V_{target} < 15$, and 30~mas for faint targets 
$15 < V_{target} < 19$. The \textit{fringe tracker} will set the coarse delay
on the target with an accuracy of 10~nm (for $V_{target} < 10$, corresponding
to a maximum offset of $\approx 0.2$ mas at 500~nm.
\item{\it Spectral dispersion:} A thin prism is inserted into the light path
after beam combination, turning the instrument into an
\textit{objective prism spectrograph}, such that the images of stars in the
focal plane are stretched out into spectra in a direction approximately
parallel to the projection of the SIM baseline on the sky. For a target at
the center of the FOV, the experimentally-measured dispersion can be modeled
with the polynomial
$x = \sum_{i=0}^6c_j\nu^j$ with $c_0 = 5.1$, $c_1 = -2.7$, $c_2 = 1.0$,
$c_3 = -0.23$, $c_4 = 0.03$, $c_5 = 0.002$, and $c_6 = 0.00005$, for $\nu$  
in units of 10$^{14}$Hz and $x$ in mm.

\item {\it Focal plane camera:} The camera has pixels of size 24 $\mu$
aligned along the direction of dispersion; however, we ignore the pixellation of
the focal plane camera in the simulations described in this paper.%

\item{\it Overall measurement precision:} The design goal is to make a single
measurement of the angular position of a target on the sky with a precision
of $\lesssim 10$ \muas.  As the \textit{fringe period}
$\lambda/B$ is typically 10 - 20 mas and the baseline length is about
$10 - 20 \times 10^6 \lambda$, the design requirement on the
single-measurement angular precision corresponds to $\approx 1/1000$
of a fringe.
\end{description}

\subsection{Estimation of Total Power}
\label{appsec:power_estimation}

In this Appendix we present the mathematical details of how we estimated the
total light from the target and field stars that lie within and just outside
the 3\arcsec\ FOV of SIM.
\begin{figure}[ht]
\epsscale{.60}
\plotone{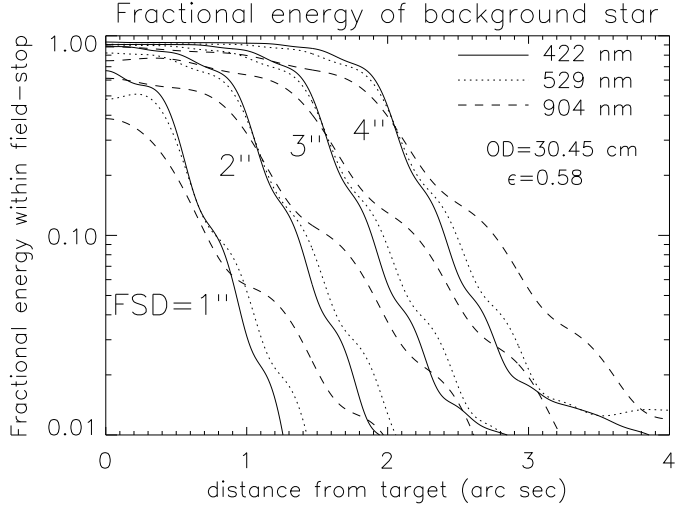}
\caption{Fraction of the energy recorded for a star located at a given
radial distance from the center of the (circular) FOV of SIM, for field
stops of 2, 3 and 4\arcsec\ diameter, and at several different wavelengths.
\label{fig:FOVresponse}
}
\end{figure}

For the target star, the total energy received by the interferometer in each
channel is multiplied by the throughput and the encircled energy within the
aperture at the mean wavelength $\bar{\lambda}$ (to account for the FOV set
by the field-stop) and integrated over the bandpass to obtain the total
power in that channel. It is then multiplied by $hc/{\bar{\lambda}}$ to
obtain the number of photons/second/channel.

In the case of the field stars, the total energy is obtained as follows:
Consider a Cartesian coordinate system centered on the target star 
which is also at the center of the FOV. Suppose that there is a field star
at ($x_0,y_0$). The energy of that part of the diffracted image that lies
within the field stop is obtained by calculating the integral of the product
of the point spread function (PSF) centered at ($x_0,y_0$) and a pill-box
function representing the field-stop. This is equivalent to integrating the
product of the on-axis PSF representing the field star with a shifted version
of the pill-box function centered at ($x_0,y_0$). The expression is:
\begin{equation}
EF_{bg\star}(x_0,y_0) = \int \int PSF(x,y) \times
\mbox{Circ}(x-x_0,y-y_0) dx dy
\end{equation}
where the ``Circ'' function is defined as:
\begin{center}
\[ \mbox{Circ($x,y$)} =\left \{ \begin{array}{ll}
            1 & \mbox{if $ x^2+y^2 < (FSD/2)^2$} \\
            0 &  \mbox{otherwise,}
           \end{array}
\right. \]
\end{center}
and $EF$ is the estimated flux, PSF is the point spread function, and FSD
is the field stop diameter. From this representation it is clear that the
energy of the field star that lies within the field stop is
the convolution (or correlation) of the on-axis PSF of the field star with
a pill-box function located at ($x_0,y_0$). It is convenient to normalize
this correlation function and obtain a fractional number which, when
multiplied by the total energy received by the interferometer from the field
star, provides the desired value. This value is further multiplied by the 
throughput of the system and converted into photons/second/channel for each
field star. Figure \ref{fig:FOVresponse} shows this overall transmission
function for several possible field stop sizes and wavelengths. We have
computed this response function at the mean wavelength of each channel.


\end{document}